\documentclass{article}
\usepackage{graphicx} 
\usepackage[utf8]{inputenc}
\usepackage[margin=2.5cm]{geometry}
\usepackage{amsmath}
\usepackage{amssymb}
\usepackage{cite}
\usepackage{authblk}
\usepackage{graphicx}
\usepackage{xcolor}
\usepackage{cancel}
\usepackage{ulem}
\usepackage{array}
\definecolor{link}{rgb}{0.0, 0.0, 0.0 }
\usepackage[bookmarks = true,
			pdfstartview = FitH,
			colorlinks = true,
			urlcolor=link,
			citecolor=link,
			linkcolor=link,
			hyperfootnotes=false]{hyperref}

\begin{document}

\title{Input-output formulation of quantum light spectroscopy}
\author[1,2]{Liwen Ko\thanks{liwen.jko@berkeley.edu}}
\author[1,2]{Robert L. Cook\thanks{rlcook@berkeley.edu}}
\author[1,2]{K. Birgitta Whaley\thanks{whaley@berkeley.edu}}
\affil[1]{Department of Chemistry, University of California, Berkeley, CA 94720, USA}
\affil[2]{Kavli Energy Nanoscience Institute at Berkeley, Berkeley, CA 94720, USA}

\maketitle
\begin{abstract}
We develop an input-output formulation of quantum light spectroscopy (QLS) by combining the input-output theory, commonly used in quantum optics, with the perturbative expansion method for nonlinear spectroscopy, commonly used in chemical physics. The use of the input-output relation provides physical intuition for how matter systems alter the photon input field. As examples, we consider linear spectroscopy, fluorescence, intra-beam $G^{(2)}$, and Hong-Ou-Mandel correlation. For certain initial photon states, we show that signal expectation values can be evaluated non-perturbatively. For general initial photon states, the signals can be computed via a perturbative expansion of the input-output relation. The input-output relation reduces the work needed to perturbatively expand the optical signal, allowing for cleaner derivations and clearer physical interpretations.
\end{abstract}

\section{Introduction}
Due to recent technological advances in the generation, manipulation, and detection of non-classical light, non-classical properties of light, such as photon antibunching \cite{wientjes2014strong, Cui_2013_quantum_imaging,lupton2021review,hettich2002nanometer,Kimble_1977}, entanglement \cite{shadbolt2012generating,edamatsu2004generation,stevenson2007biphoton,Kwiat_1995,sultanov2024tunable}, squeezing \cite{wu1987squeezed, aggarwal2020room,ourjoumtsev2011observation,Wu_1986_squeezed,safavi2013squeezed}, or Hong-Ou-Mandel interference \cite{Kalashnikov_2017_HOM, HOM_1987, lettow2010quantum, Sipahigil_2014,deng2019quantum}, have become commonly observed in the laboratory. Researchers have exploited, or have proposed to exploit, these non-classical properties in various applications, such as quantum computation \cite{zhong2020quantum, slussarenko2019photonic,obrien2009photonic}, enhancing the detection of gravitational waves \cite{caves1981quantum,aasi2013enhanced}, imaging \cite{abouraddy2002quantum, lemos2014quantum, Cui_2013_quantum_imaging}, x-ray diffraction \cite{asban2019quantum}, optical lithography \cite{Boto_2000_lithography}, and molecular spectroscopy \cite{Mukamel_Rev_Mod_Phys, Schlawin_2017_tutorial}.
\par
In particular, quantum light spectroscopy (QLS) aims to enhance or surpass traditional spectroscopy experiments that use classical-like coherent state laser light. This can be done either by probing the matter systems with non-classical light or by detecting the non-classical properties of the emitted light. For example, entangled photon pairs can be used to excite doubly excited states with high spectral specificity \cite{Schlawin_2012_double_excitation}, to act as an analog of two-dimensional electronic spectroscopy \cite{ishizaki2020probing}, or to modify the probability of two-photon absorption \cite{schlawin2018entangled, Raymer_2021_tutorial}.
Squeezed light can be used as a low-noise light source for spectroscopy \cite{grangier1987squeezed,polzik1992spectroscopy}.
The second order coherence function $g^{(2)}(\tau)$ of the fluorescent light contains information about the transient dynamics in the matter system \cite{lupton2021review}, and it has also been claimed to be a witness for quantum coherence in the matter system \cite{munoz_g2_2020,Holdaway_2018, nation2024photoncorrelationtimeasymmetrydynamical,nation2024twocolourphotoncorrelationsprobe}.
\par
A popular theoretical framework to study the interaction between quantum light and matter is the input-output theory \cite{Gardiner1985}, which is commonly used in the quantum optics community to study atoms or cavities interacting with one-dimensional propagating photon fields \cite{gardiner_zoller_quantum_noise}. The input-output theory provides an exact expression, known as the input-output relation, that relates the input field to the output field. It also provides formally exact results for the master equation and the Heisenberg equation of motion for the system operators, known as the Heisenberg-Langevin equation \cite{Gardiner1985,Combes_2017_review}. 
However, the exact formal results often need to be further simplified in a case-by-case basis, depending on the type of the input field state. For example, the master equations for a matter system under the excitation of a coherent state \cite{wiseman_milburn_2009_book}, a thermal state \cite{Gardiner1985}, a squeezed state \cite{Gardiner1985}, and an $m$-photon Fock state \cite{Baragiola_2012} take different forms.
\par
Another theoretical framework to treat the fully quantum light-matter interaction is the perturbative approach \cite{Mukamel_Rev_Mod_Phys, Schlawin_2017_tutorial}, which has been developed in the chemical physics community as a quantum optical generalization of the perturbative formalism used for describing classical nonlinear spectroscopy \cite{Mukamel_book}. In classical nonlinear spectroscopy, one performs a perturbative expansion of the matter state under the interaction with a classical photon field, while in the generalization to QLS, one performs a perturbative expansion of the combined matter+field state \cite{Mukamel_Rev_Mod_Phys, Schlawin_2017_tutorial}. The perturbative framework is restricted to the weak light-matter coupling regime, but it provides a general method to treat the effects from all types of input photon states. This is because the effects from different orders of light-matter interaction are manifested in the photon field correlation functions of different orders, which can be evaluated for all types of input field states.
\par
\begin{figure}
    \centering
    \includegraphics[scale=0.58]{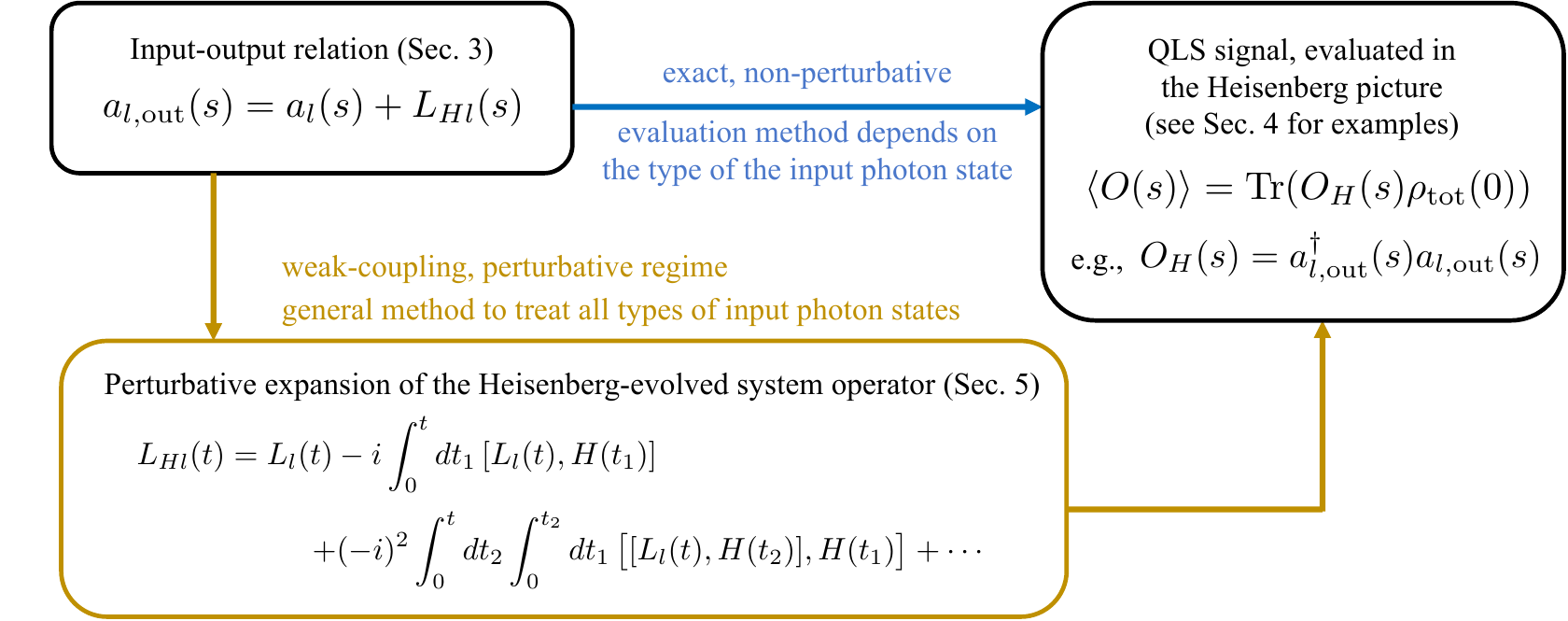}
    \caption{The input-output formalism for QLS provides a unified framework to treat the quantum light-matter interaction in both the non-perturbative and perturbative regimes. The input-output relation, discussed in Sec. \ref{Sec:input_output_relation}, is an exact result under the Hamiltonian of Eq. (\ref{Eq:H_int_main}) or (\ref{Eq:H_int_main_many_molecules}). Examples of using the input-output relation to analyze various QLS signal observables $O$ are discussed in Sec. \ref{Sec:spectroscopy_using_IO_examples}. In the weak coupling regime, the exact input-output relation can be expanded perturbatively in the Heisenberg picture. This is discussed in Sec. \ref{Sec:perturbative_expansion_of_output_field}. }
    \label{fig:concept_map}
\end{figure}

In this paper, we develop an input-output formalism for QLS, combining the input-output theory with the perturbative expansion method for nonlinear spectroscopy.
This new formalism provides an intuitive and unified description for various types of QLS experiments in both the perturbative and non-perturbative regimes. In the general non-perturbative regime, the input-output relation is an exact result that connects the input field to the output field. In the perturbative regime, we demonstrate how to expand the input-output relation perturbatively, allowing us to treat different types of input light generally. Fig. (\ref{fig:concept_map}) illustrates how our input-output formalism unifies the description of QLS in both the perturbative and non-perturbative regimes.
\par
There are some important differences between our input-output formalism for QLS and the conventional perturbative formalism for QLS \cite{Mukamel_Rev_Mod_Phys, Schlawin_2017_tutorial}. 
First, when analyzing the optical signal, we work in the Heisenberg picture, which is analogous to how classical electromagnetic field evolves in time. This is in contrast to the conventional perturbative approach for QLS, which works in the interaction picture.
Second, we distinguish between the real time and the retarded time, while the conventional formalism usually only keeps track of the real time. The real time and retarded time together provide information on both the space and time dependence of the photon field.  
The lack of distinction between the real time and the retarded time in the conventional approach could lead to an ambiguity in the integration bounds in the perturbative expansion. This issue is explained in detail in Sec. \ref{Sec:conventional_perturbative_approach}.
Third, we make use of the input-output relation in the analysis of the optical signal. Since this is an exact relation, it provides a way to analyze the signal non-perturbatively. Furthermore, the input-output relation, which states that the output field is equal to the input field plus some Heisenberg-evolved matter operator (see Eq. (\ref{Eq:input_output_relation}) or (\ref{Eq:input_output_relation_single_molecule})), has a close analogy in classical nonlinear spectroscopy. In classical nonlinear spectroscopy, the output signal electric field is expressed as the input electric field plus the electric field generated by the matter polarization (i.e., dipole moment per unit volume) \cite{Mukamel_book}. The input-output relation is therefore a quantum mechanical generalization of this classical statement. 
The added intuition from our input-output formalism for QLS has led us to discover an equivalence between a class of QLS and a class of classical light spectroscopy \cite{Ko_2023}. 
\par
We now give an overview of the content in this paper.
Sec. \ref{Sec:matter_field_interaction} expands upon our previous work \cite{Ko_2022} to show how to rigorously describe the interaction between a matter system and the 3-dimensional photon field as the interaction between the matter system and a finite number of 1-dimensional photon field spatial modes. 
Sec. \ref{Sec:input_output_relation} reviews the input-output relation, emphasizing the distinction between real time and retarded time.
In Sec. \ref{Sec:spectroscopy_using_IO_examples}, we examine several common spectroscopic setups and demonstrate how to use the input-output relation to analyze different kinds of optical signals. Exact, non-perturbative results can be obtained for special input states. 
Sec. \ref{Sec:perturbative_expansion_of_output_field} describes the perturbative expansion of the Heisenberg picture input-output relation and compares with the conventional perturbative method in the interaction picture. A normal-ordered expansion method is described in Sec. \ref{Sec:normal_ordered_perturbative_expansion_operator}.

\section{Matter -- photon field interaction Hamiltonian and field observables}
\label{Sec:matter_field_interaction}
We take the dipole -- electric field Hamiltonian as the fundamental Hamiltonian for light-matter interaction. The electric field is a 3-dimensional field (3 space dimensions, not counting the time dimension), i.e., its degrees of freedom are indexed by the 3-dimensional real space or wavevector space coordinates. However, the conventional perturbative approach for QLS \cite{Mukamel_Rev_Mod_Phys, Schlawin_2017_tutorial} and the input-output formalism \cite{Gardiner1985,gardiner_zoller_quantum_noise} treat the photon fields as 1-dimensional fields. The 1-dimensional fields in these theories are usually considered as a model for plane wave photons or photons confined in an infinite cylinder with some cross section area $A$. This picture is correct for photons in waveguides, but it is not a satisfactory description of photons in 3-dimensional space, as it does not properly account for all 3-dimensional degrees of freedom. To bridge this gap in the different descriptions of the photon field, we have shown in our earlier paper \cite{Ko_2022} how to rigorously decompose the 3-dimensional photon field into a countably infinite number of 1-dimensional fields. The electric field operator, however, can be expressed in terms of a finite number of these 1-dimensional fields. Therefore, the remaining infinitely many 1-dimensional fields decouple from the matter system and evolve freely. The general procedure for this reduction from a 3-dimensional field into 1-dimensional fields is summarized in Appendix \ref{app:photon_as_1D_fields}. In this paper, we use specifically the small solid angle decomposition scheme to express the electric field in terms of a finite number of 1-dimensional small solid angle spatial modes. 
We extend our presentation of the small solid angle decomposition in \cite{Ko_2022} to study the spatial properties of these small solid angle modes here. We show that this decomposition to a finite number of 1-dimensional fields is valid only within an interaction region in real space. Under the narrow-band approximation, the small solid angle modes resemble Gaussian paraxial modes in real space and time. 

\subsection{Total Hamiltonian with the dipole -- electric field interaction}
\label{sec:dipole_electric_field_Hamiltonian}
We take the combined matter system plus photon field Hamiltonian to be 
\begin{equation}
    H_{\text{sys+field}} = H_{\text{sys}} + H_{\text{field}} + H_{\text{coup}},
\label{Eq:full_Hamiltonian_Schrodinger_picture}
\end{equation}
where $H_{\text{sys}}$, $H_{\text{field}}$, and $H_{\text{coup}}$ are the Hamiltonians for the matter system, the photon field, and the coupling between the matter system and the photon field \cite{Loudon_2000_book}. In realistic models of molecular systems, $H_{\text{sys}}$ would typically contain both the electronic and nuclear degrees of freedom. $H_{\text{field}}$ is
\begin{equation}
    H_{\text{field}} = \int d^3 k \sum_\lambda \hbar c|\mathbf{k}| a^\dagger_{\mathbf{k},\lambda} a_{\mathbf{k},\lambda},
\label{Eq:H_field_1}
\end{equation}
where $\mathbf{k}$ is the 3-dimensional wavevector, and $\lambda$ indexes the two possible polarizations corresponding to each $\mathbf{k}$. $\hbar$ and $c$ are the reduced Planck constant and the speed of light, respectively. $a_{\mathbf{k},\lambda}$ and $a^\dagger_{\mathbf{k},\lambda}$ are the bosonic annihilation and creation operators for the field mode $(\mathbf{k},\lambda)$, and they satisfy the bosonic commutation relations:
\begin{subequations}
\begin{equation}
    [a_{\mathbf{k},\lambda}, a_{\mathbf{k'},\lambda'}]=[a^\dagger_{\mathbf{k},\lambda}, a^\dagger_{\mathbf{k'},\lambda'}]=0
\end{equation}
and
\begin{equation}
    [a_{\mathbf{k},\lambda}, a^\dagger_{\mathbf{k'},\lambda'}]=\delta(\mathbf{k}-\mathbf{k'})\delta_{\lambda,\lambda'}.
\end{equation}
\label{Eq:3D_commutation_relations}
\end{subequations}
\par For atomic or molecular systems, the size of the matter system is usually small compare to the wavelength of light it interacts with, so that the dipole approximation holds, and the coupling Hamiltonian $H_{\text{coup}}$ takes the dipole -- electric field form $H_{\text{coup}} = -\mathbf{d}\cdot\mathbf{E}(\mathbf{x})$, where $\mathbf{x}$ is the position of the matter system.
The electric field operator $\mathbf{E}(\mathbf{x})$, expressed in terms of $a_{\mathbf{k},\lambda}$, is
\begin{equation}
    \mathbf{E}(\mathbf{x}) = \int \frac{d^3 \mathbf{k}}{(2\pi)^{3/2}} \sum_\lambda \sqrt{\frac{\hbar\omega}{2\epsilon_0}} (ia_{\mathbf{k},\lambda}e^{i\mathbf{k}\cdot\mathbf{x}}-ia^\dagger_{\mathbf{k},\lambda}e^{-i\mathbf{k}\cdot\mathbf{x}})\hat{\mathbf{e}}_{\mathbf{k},\lambda},
\label{Eq:E_field_1}
\end{equation}
where $\epsilon_0$ is the permittivity of free space and $\hat{\mathbf{e}}_{\mathbf{k},\lambda}$ is the unit vector in the direction of the polarization in the field mode $(\mathbf{k},\lambda)$. 

\par We write the electric field $\mathbf{E}(\mathbf{x})$ as the sum $\mathbf{E}(\mathbf{x})=\mathbf{E}^{(+)}(\mathbf{x})+\mathbf{E}^{(-)}(\mathbf{x})$. The positive frequency component $\mathbf{E}^{(+)}(\mathbf{x})$ is the part that contains annihilation operators $a_{\mathbf{k},\lambda}$. The negative frequency component $\mathbf{E}^{(-)}(\mathbf{x})$ is the part that contains creation operators $a^\dagger_{\mathbf{k},\lambda}$. $\mathbf{E}^{(+)}(\mathbf{x})$ and $\mathbf{E}^{(-)}(\mathbf{x})$ are Hermitian conjugates of each other. We also decompose the matter dipole operator $\mathbf{d}$ into an excitation component $\mathbf{d}^{(+)}$ that creates an excitation in the matter and a de-excitation component $\mathbf{d}^{(-)}$ that removes an excitation in the matter, so that $\mathbf{d}=\mathbf{d}^{(+)} + \mathbf{d}^{(-)}$ and $\mathbf{d}^{(+)} = \mathbf{d}^{(-)\dagger}$. Under the rotating wave approximation, the total number of excitations is conserved, and the interaction term $H_{\text{coup}}$ becomes
\begin{equation}
    H_{\text{coup}} = -\mathbf{d}^{(+)}\cdot\mathbf{E}^{(+)}(\mathbf{x})-\mathbf{d}^{(-)}\cdot\mathbf{E}^{(-)}(\mathbf{x}).
\label{Eq:H_coup_single_molecule}
\end{equation}

\par In spectroscopy experiments, the material sample usually consists of a large number of matter systems (e.g., atoms or molecules) that do not interact with one another. We will hereafter refer to the matter systems as molecules. In this case, the interaction term is a sum over all molecules, i.e., 
\begin{equation}
	H_{\text{coup}} = \sum^N_{j=1} - \mathbf{d}^{(+)}_j\cdot\mathbf{E}^{(+)}(\mathbf{x}_j) - \mathbf{d}^{(-)}_j\cdot\mathbf{E}^{(-)}(\mathbf{x}_j),
\label{Eq:H_coup_many_molecules}
\end{equation}
where $j$ indexes the $N$ non-interacting molecules. $\mathbf{d}_j$ is the dipole operator for the $j$-th molecule, and $\mathbf{x}_j$ is the position of the $j$-th molecule.

\subsection{Using small solid angle decomposition to express the 3-dimensional electric field as a finite number of 1-dimensional fields}
\label{sec:3d_to_1d}

The photon field is a 3-dimensional field, since the field operator $a_{\mathbf{k},\lambda}$ is indexed by the 3-dimensional wavevector $\mathbf{k}$. The input-output formalism, however, works with 1-dimensional fields. A rigorous conversion from the 3-dimensional photon field $a_{\mathbf{k},\lambda}$ into 1-dimensional fields $a_l(\omega)$ is provided in \cite{Ko_2022} and is summarized in Appendix \ref{app:photon_as_1D_fields}. Here, we focus specifically on the small solid angle decomposition scheme and study the properties of the small solid angle modes. The small solid angle decomposition is a natural description for QLS experiments that use light in different spatial directions to excite, probe, or detect the matter sample.

\par
\begin{figure}
    \centering
    \includegraphics[scale=0.5]{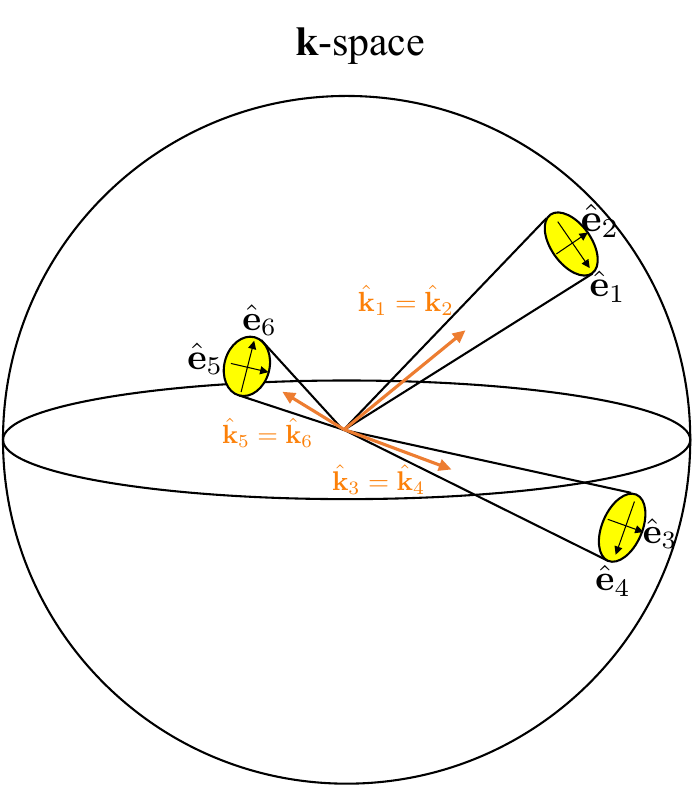}
    \caption{Schematics of the small solid angle decomposition. The $4\pi$ solid angle in $\mathbf{k}$-space is partitioned into $M$ small solid angle sections (colored yellow). Each small solid angle section is a cone in $\mathbf{k}$-space and contains two spatial modes, corresponding to the two polarizations.}
    \label{fig:small_solid_angle}
\end{figure}

As discussed in Appendix \ref{app:photon_as_1D_fields}, we partition the $4\pi$ solid angle of the orientation of $\mathbf{k}$ into $M$ small solid angle sections. Each small solid angle section contains two modes, corresponding to the two possible polarizations within each small solid angle section (see Fig. (\ref{fig:small_solid_angle})). Therefore, the electric field operator is described in terms of $2M$ (a finite number) 1-dimensional fields. Combining Eq. (\ref{Eq:a_l_definition}) and Eq. (\ref{Eq:g_small_solid_angle_def}), the 1-dimensional field operator $a_l(\omega)$ is defined in terms of the 3-dimensional field operator $a_{\mathbf{k},\lambda}$ as
\begin{equation}
    a_l(\omega) = \sqrt{\frac{\omega^2}{c^3\Delta\Omega_l}}\int_{\Omega_l} d\Omega \, a_{|\mathbf{k}|=\omega/c, \Omega,\lambda},
\label{Eq:small_angle_1D_field_definition}
\end{equation}
where $l$ indexes the $2M$ small solid angle modes. In Eq. (\ref{Eq:small_angle_1D_field_definition}), we write the 3-dimensional field operator $a_{\mathbf{k},\lambda}$ as $a_{|\mathbf{k}|,\Omega, \lambda}$, where $|\mathbf{k}|$ is the magnitude of $\mathbf{k}$, and $\Omega$ is the orientation of $\mathbf{k}$. In the integrand $a_{|\mathbf{k}|, \Omega,\lambda}$, the polarization $\lambda$ is restricted to be the same as the polarization of the $l$-th mode, and $|\mathbf{k}|$ is restricted to be equal to $\omega/c$. The integral $\int_{\Omega_l} d\Omega$ means that the orientation integral is performed on the small solid angle section $\Omega_l$ subtended by the $l$-th mode. $\Delta\Omega_l$ is the area of the $l$-th small solid angle section, expressed in unit of steradian, or $\text{rad}^2$. $a_l(\omega)$ satisfies the standard bosonic commutation relations (see Eq. (\ref{Eq:1D_boson_commutation_relations})).

\par
Under the small solid angle decomposition, the electric field operator at position $\mathbf{x}=\mathbf{0}$ is
\begin{equation}
    \mathbf{E}(\mathbf{0})=\sum_{l=1}^{2M}\int^\infty_0 d\omega \sqrt{\frac{\hbar\omega^3\Delta\Omega_l}{16\pi^3\epsilon_0 c^3}}(ia_l(\omega)-ia^\dagger_l(\omega))\hat{\mathbf{e}}_l,
\label{Eq:E_small_angle_x0}
\end{equation}
(see Eq. (\ref{Eq:E_field_finite_1D_app})), where $\hat{\mathbf{e}}_l$ is the unit vector of the polarization in mode $l$. If the solid angle sections are small enough, then the small variation of the polarization vectors $\hat{\mathbf{e}}_{\mathbf{k},\lambda}$ within a solid angle section can be ignored. Therefore we can define two constant polarization vectors $\hat{\mathbf{e}}_l$ for each solid angle section. Likewise, if the solid angle sections are small enough, we can use one representative wavevector direction $\hat{\mathbf{k}}_l$ (the unit vector of $\mathbf{k}_l$) in each solid angle section to approximate all wavevector directions $\hat{\mathbf{k}}$ in that solid angle section. 
\par
Away from the origin $\mathbf{x} = \mathbf{0}$, but close to the origin, we approximate the electric field by adding spatial phase factors to Eq. (\ref{Eq:E_small_angle_x0}). The electric field becomes
\begin{equation}
    \mathbf{E}(\mathbf{x}) \approx \sum_{l=1}^{2M}\int^\infty_0 d\omega \sqrt{\frac{\hbar\omega^3\Delta\Omega_l}{16\pi^3\epsilon_0 c^3}}(ia_l(\omega)e^{i\omega \hat{\mathbf{k}}_l \cdot \mathbf{x}/c}-ia^\dagger_l(\omega)e^{-i\omega \hat{\mathbf{k}}_l \cdot \mathbf{x}/c})\hat{\mathbf{e}}_l.
\label{Eq:E_small_angle}
\end{equation}
For this approximation to be valid, we require that the phase factor $e^{i\omega \hat{\mathbf{k}}_l \cdot \mathbf{x}/c}$ of the small solid angle mode $l$ accurately represent all other phase factors $e^{i\omega\hat{\mathbf{k}}\cdot\mathbf{x}/c}$ (with the same frequency $\omega$) in the same small solid angle mode $l$. This is true if $(\omega\hat{\mathbf{k}}_l/c - \omega\hat{\mathbf{k}}/c )\cdot \mathbf{x} \ll 1$, or equivalently, $\Delta\theta |\mathbf{k}| \cdot \mathbf{x} \ll 1$, where $|\mathbf{k}| = \omega/c$ is the magnitude of $\mathbf{k}$, and $\Delta\theta$ is the angular width of the small solid angle section. 
We will see that $\Delta \theta$ corresponds to the beam divergence angle in real space. 
Therefore, Eq. (\ref{Eq:E_small_angle}) is valid in the region $\mathbf{x}\ll 1/(\Delta\theta|\mathbf{k}|)$. The quantity $1/(\Delta\theta|\mathbf{k}|)$ defines the length scale of the interaction region, where the matter system interacts with photons in the small solid angle spatial modes.
For example, if the beam divergence angles in an experiment are on the order of $\Delta\theta\sim 10^{-3}\,\text{rad}$ and if the characteristic wavelengths of the spatial modes are centered at $800\,\text{nm}$ (equal to $2\pi/|\mathbf{k}|$), then the interaction region for the sample has a characteristic length of $1/(\Delta\theta|\mathbf{k}|) \approx 100 \,\mu\text{m}$.
\par
Whereas the electric field is expressed as a sum of a finite number of 1-dimensional field modes, the field Hamiltonian is expressed as a sum over a countably infinite number of 1-dimensional field modes
\begin{equation}
    H_{\text{field}} = \sum_{l=1}^\infty \int^\infty_0 d\omega\, \hbar\omega a^\dagger_l(\omega)a_l(\omega).
\label{Eq:H_field_1D_general}
\end{equation}
(see Appendix \ref{app:photon_as_1D_fields}). The first $2M$ modes are defined explicitly in Eq. (\ref{Eq:small_angle_1D_field_definition}). The remaining infinitely many modes can be constructed in principle, but we do not need to define them explicitly.
This is because only the first $2M$ modes appear in the electric field operator $\mathbf{E}(\mathbf{x})$. The remaining infinitely many field modes do not appear in $\mathbf{E}(\mathbf{x})$, and therefore do not appear in the light-matter coupling term $H_{\text{coup}}$, so they decouple from the matter system and evolve freely. If one is only interested in the finite number of 1-dimensional field modes that couple to the matter system, then one can ignore the remaining infinite number of 1-dimensional fields. For most spectroscopy experiments near the visible regime, the remaining infinitely many fields will be in the vacuum state, and stay in the vacuum state under free evolution.

\subsubsection{Examining the small solid angle modes in real space and time}
\begin{figure}
    \centering
    \includegraphics[scale=0.5]{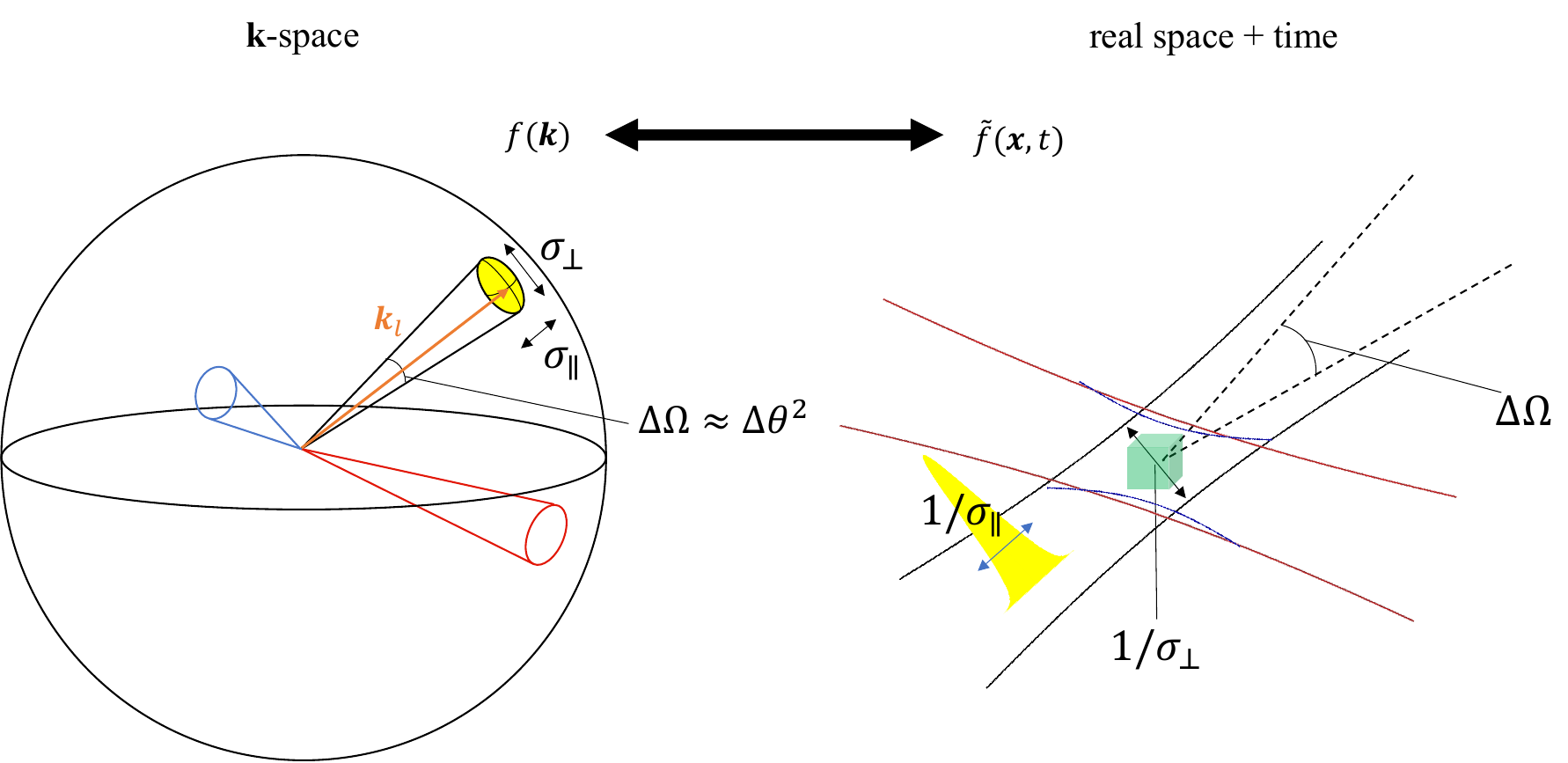}
    \caption{Three small solid angle modes, colored black, red, and blue, are represented in the $\mathbf{k}$-space and in the real space, under the narrow-band approximation. Under the narrow-band approximation, the mode function $f(\mathbf{k})$ of a small solid angle mode is concentrated in a small region, with a transversal width of $\sigma_\perp$ and a longitudinal width of $\sigma_\parallel$. The transversal width $\sigma_\perp$ is related to the angular width $\Delta \theta$ by $\sigma_\perp\sim |\mathbf{k}_l|\Delta\theta$. In real space, the small solid angle mode $\Tilde{f}(\mathbf{x},t)$ resembles a Gaussian paraxial mode. At $t=0$, the waist has a transversal area on the order of $1/\sigma_\perp^2$. As $t\rightarrow\infty$, the mode subtends a solid angle area of $\Delta\Omega$ asymptotically. The paraxial pulse has a longitudinal width of $1/\sigma_\parallel$ in real space. The green cube is in the overlapping region of different spatial modes, thus representing an interaction region where multiple spatial modes can interact with the matter sample in a QLS experiment. The interaction region has a characteristic length of $1/\sigma_\perp\sim 1/(\Delta\theta|\mathbf{k}_l|)$.}
    \label{fig:paraxial}
\end{figure}

The small solid angle modes, defined in the $\mathbf{k}$-space, can also be interpreted intuitively in real space by invoking the narrow-band approximation. The narrow-band approximation makes use of the fact that, in spectroscopy experiments near the visible regime, the matter system only interacts significantly with a narrow band of frequency around some characteristic frequency $\omega_0$ of the matter system (i.e., $\omega\in (\omega_0-\Delta\omega, \omega_0+\Delta\omega)$, where $\Delta\omega \ll \omega_0$). To model the small solid angle mode under the narrow-band approximation, consider a small region in $\mathbf{k}$-space in the $l$-th small solid angle mode, centered around $\mathbf{k}_l=\omega_0/c \hat{\mathbf{k}}_l$ (see Fig. (\ref{fig:paraxial})). We denote the transversal width of the small region to be $\sigma_\perp$, so that $\sigma_\perp\sim \Delta\theta|\mathbf{k}_l|$. The angular width $\Delta\theta$ is related to the small solid angle area $\Delta\Omega$ by $\Delta\Omega\sim\Delta\theta^2$. The longitudinal width of the region is denoted as $\sigma_\parallel$, and it is related to the frequency bandwidth by $\sigma_\parallel = \Delta\omega/c$. 
\par
To understand the behavior of the photon field due to this small region in $\mathbf{k}$-space, let us construct a function $f(\mathbf{k})$, whose value is nonzero and slowly-varying when $\mathbf{k}$ is in the small region, and we let $f(\mathbf{k})$ drop to $0$ quickly away from the small region. In real space and time, the mode $f(\mathbf{k})$ is given by
\begin{equation}
    \Tilde{f}(\mathbf{x},t)=\int d^3\mathbf{k} \, f(\mathbf{k}) e^{i\mathbf{k}\cdot \mathbf{x}} e^{-ic|\mathbf{k}|t}.
\label{Eq:f_real_space_time}
\end{equation}
We show in Appendix \ref{app:small_solid_angle_real_space} that $\Tilde{f}(\mathbf{x},t)$ resembles a Gaussian paraxial mode in real space and time. As illustrated in Fig. (\ref{fig:paraxial}), at time $t=0$, the field amplitude is centered at $\mathbf{x}=\mathbf{0}$, and has a cross sectional area on the order of $1/\sigma_\perp^2$. As $t\rightarrow \infty$, the field amplitude spreads out into a cone with a solid angle area of $\Delta\Omega$, same as the solid angle area in the $\mathbf{k}$-space. The paraxial pulse has a longitudinal width of $\sim 1/\sigma_\parallel$. 
\par
Because the field amplitude has a cross sectional area of $\sim 1/\sigma_\perp^2$ near the origin, different spatial modes overlap in a region with a characteristic length scale $1/\sigma_\perp$ near the origin (see right side of Fig. (\ref{fig:paraxial})). This defines an interaction region where multiple spatial modes can interact with the matter sample in a QLS experiment. The characteristic length scale $1/\sigma_\perp\sim 1/(\Delta\theta|\mathbf{k}_l|)$ is consistent with our previous discussion in the $\mathbf{k}$-space.

\subsection{The interaction picture Hamiltonian}
\label{sec:interaction_pic_Hamiltonian}
Consider a molecule located at position $\mathbf{x}$. Using the electric field expression in Eq. (\ref{Eq:E_small_angle}), the dipole -- electric field coupling Hamiltonian in Eq. (\ref{Eq:H_coup_single_molecule}) becomes
\begin{equation}
    H_{\text{coup}} = \sum_{l=1}^{2M} \int^\infty_{-\infty} \frac{d\omega}{\sqrt{2\pi}} -ia_l(\omega)L_l^\dagger e^{i\omega \hat{\mathbf{k}}_l \cdot \mathbf{x}/c} +ia^\dagger_l(\omega)L_l e^{-i\omega \hat{\mathbf{k}}_l \cdot \mathbf{x}/c}.
\label{Eq:H_int_1}
\end{equation}
We have combined the prefactor in the electric field with the dipole operator $\mathbf{d}$ into a scaled dipole operator $L_l$, defined by
\begin{equation}
    L_l = \sqrt{\frac{\hbar\omega_0^3\Delta\Omega_l}{8\pi^2 \epsilon_0 c^3}}\mathbf{d}_-\cdot\hat{\mathbf{e}}_l.
\label{Eq:L_definition}
\end{equation}
Note that $L_l$ has a physical dimension of $[1/\sqrt{\text{time}}]$.
In Eq. (\ref{Eq:H_int_1}), we have used the narrow-band approximation (also known as the white noise approximation) to replace the variable $\omega$ in the prefactor of Eq. (\ref{Eq:E_small_angle}) with a fixed characteristic frequency $\omega_0$. Under the narrow-band approximation, the integration range $(0,\infty)$ in Eq. (\ref{Eq:E_small_angle}) is extended to $(-\infty,\infty)$ in Eq. (\ref{Eq:H_int_1}), since only the photons in the narrow frequency window $(\omega_0-\Delta\omega, \omega_0+\Delta\omega)$ affect the overall dynamics significantly. 
The factor of $1/\sqrt{2\pi}$ in Eq. (\ref{Eq:H_int_1}) is separated out for later convenience, as we will see below. 
Eq. (\ref{Eq:H_int_1}) is generalized to the case of many molecules by summing over all molecules, i.e., 
\begin{equation}
    H_{\text{coup}} = \sum_{l=1}^{2M} \sum_{j=1}^N \int^\infty_{-\infty} \frac{d\omega}{\sqrt{2\pi}} -ia_l(\omega)L_{l,j}^\dagger e^{i\omega \hat{\mathbf{k}}_l \cdot \mathbf{x}_j/c} +ia^\dagger_l(\omega)L_{l,j} e^{-i\omega \hat{\mathbf{k}}_l \cdot \mathbf{x}_j/c},
\label{Eq:H_int_1_many_molecules}
\end{equation}
where $L_{l,j}$ is the scaled dipole operator that couples the $j$-th molecule to the $l$-th spatial mode, and $\mathbf{x}_j$ is the position of the $j$-th molecule. 
\par
Now, we transform into an interaction picture by writing the total Hamiltonian as
\begin{equation}
    H_{\text{sys+field}} = H_0 + H_{\text{coup}},
\end{equation}
where
\begin{equation}
    H_0=H_{\text{sys}}+H_{\text{field}}.
\end{equation}
A general Schrodinger picture operator $A$ transforms into $A(t) = e^{iH_0(t-t_0)}Ae^{-iH_0 (t-t_0)}$ in the interaction picture (setting $\hbar=1$ from now on). In the expression for $A(t)$, $t_0$ is the initial time with respect to which we define the interaction picture, and $t_0$ will be set to $0$ from now on. In particular, the interaction picture Hamiltonian is
\begin{equation}
    H(t) = e^{iH_0 t}H_{\text{coup}}e^{-iH_0 t}.
\label{Eq:H_interaction_picture}
\end{equation}
For notational simplicity, we have dropped the subscript ``coup" and simply denote the interaction picture Hamiltonian as $H(t)$, since this is the main Hamiltonian we will work with.
We define the interaction picture time evolution operator $U(t)$ as the solution of the interaction picture Schrodinger equation (in operator form)
\begin{equation}
    \frac{dU(t)}{dt} = -iH(t)U(t)
\label{Eq:Schrodinger_eqn_1}
\end{equation}
with the initial condition $U(0)=1$. The interaction picture time evolution operator $U(t)$ is then related to the Schrodinger picture time evolution operator $e^{-iH_{\text{sys+field}}t}$ by
\begin{equation}
    e^{-iH_{\text{sys+field}}t} = e^{-iH_0 t}U(t).
\label{Eq:interaction_U_to_schrodinger_U}
\end{equation}
One can check that Eq. (\ref{Eq:interaction_U_to_schrodinger_U}) is indeed correct by taking the time derivative on both sides. Using Eq. (\ref{Eq:interaction_U_to_schrodinger_U}), one can show that a Heisenberg picture operator $A_H(t)=e^{iH_{\text{sys+field}}t}Ae^{-iH_{\text{sys+field}}t}$ is related to the interaction picture operator $A(t)$ by 
\begin{equation}
    A_H(t) = U^\dagger(t) A(t) U(t).
\label{Eq:interation_to_Heisenberg_general}
\end{equation}

\par
To obtain an explicit expression for the interaction picture Hamiltonian $H(t)$ (Eq. (\ref{Eq:H_interaction_picture})), we note that the Schrodinger picture operator $a_l(\omega) e^{i\omega \hat{\mathbf{k}}_l \cdot \mathbf{x}/c}$ in Eq. (\ref{Eq:H_int_1}) becomes 
\begin{equation}
    e^{iH_0t}a_l(\omega) e^{i\omega \hat{\mathbf{k}}_l \cdot \mathbf{x}/c}e^{-iH_0t} = a_l(\omega)e^{-i\omega (t-\hat{\mathbf{k}}_l\cdot\mathbf{x}/c)}
\label{Eq:interaction_pic_al_omega}
\end{equation}
in the interaction picture. 
If we define the retarded time in the paraxial mode $l$ as
\begin{equation}
    s_{l}(t,\mathbf{x}) = t-\frac{\hat{\mathbf{k}}_l\cdot\mathbf{x}}{c},
\label{Eq:retarded_time_def}
\end{equation}
then we can re-write the operator in Eq. (\ref{Eq:interaction_pic_al_omega}) as $a_l(\omega)e^{-i\omega s_{l}}$.
As time $t$ advances, the plane corresponding to a fixed retarded time $s$ (e.g., $s=2$ in Fig. (\ref{fig:retarded_time})) moves at the speed of light in the direction of $\hat{\mathbf{k}}_l$. 
\par
We now define the retarded time -- dependent field operator (or sometimes known as the white noise operator)
\begin{equation}
    a_l(s) = \int^\infty_{-\infty} \frac{d\omega}{\sqrt{2\pi}} \, a_l(\omega)e^{-i\omega s}
\label{Eq:a_Fourier_relation}
\end{equation}
as the Fourier transform of the frequency-dependent field operator. An important consequence of the definition of $a_l(s)$ is that they satisfy the bosonic commutation relations: $[a_l(s),a_{l'}(s')]=[a^\dagger_l(s),a^\dagger_{l'}(s')]=0$ and $[a_l(s),a^\dagger_{l'}(s')]=\delta(s-s')\delta_{l,l'}$. Since $a_l(\omega)$ has a physical dimension of [$1/\sqrt{\text{frequency}}$] (see Eq. (\ref{Eq:a_l_definition})), by Eq. (\ref{Eq:a_Fourier_relation}), $a_l(s)$ has a physical dimension of [$1/\sqrt{\text{time}}$]. The single-molecule $H_{\text{coup}}$ in the interaction picture now becomes
\begin{equation}
    H(t)=\sum_{l=1}^{2M} -ia_l(s_l)L^\dagger_l(t) + ia^\dagger_l(s_l)L_l(t)
\label{Eq:H_int_main}
\end{equation}
(see Eqs. (\ref{Eq:H_int_1}) and (\ref{Eq:H_interaction_picture})).
$L_l(t)$ is the scaled dipole operator $L_l$ (see Eq. (\ref{Eq:L_definition})) in the interaction picture, and it is equal to $e^{iH_{\text{sys}}t}L_l e^{-iH_{\text{sys}}t}$. We have also suppressed the dependence on $t$ and $\mathbf{x}$ in $s_l(t,\mathbf{x})$ for notation simplicity, since the value of $t$ is clear from the context and we are only considering one molecule located at position $\mathbf{x}$. For the case of $N$ molecules, the interaction picture $H_{\text{coup}}$ becomes
\begin{equation}
    H(t) = \sum_{l=1}^{2M}\sum_{j=1}^N -ia_l\big(s_l(t,\mathbf{x}_j)\big)L^\dagger_{l,j}(t) + ia^\dagger_l\big(s_l(t,\mathbf{x}_j)\big)L_{l,j}(t).
\label{Eq:H_int_main_many_molecules}
\end{equation}
(see Eqs. (\ref{Eq:H_int_1_many_molecules}) and (\ref{Eq:H_interaction_picture})).

\subsection{Photon field observables}
\label{sec:photon_field_observables}
\begin{figure}
    \centering
    \includegraphics[scale=0.5]{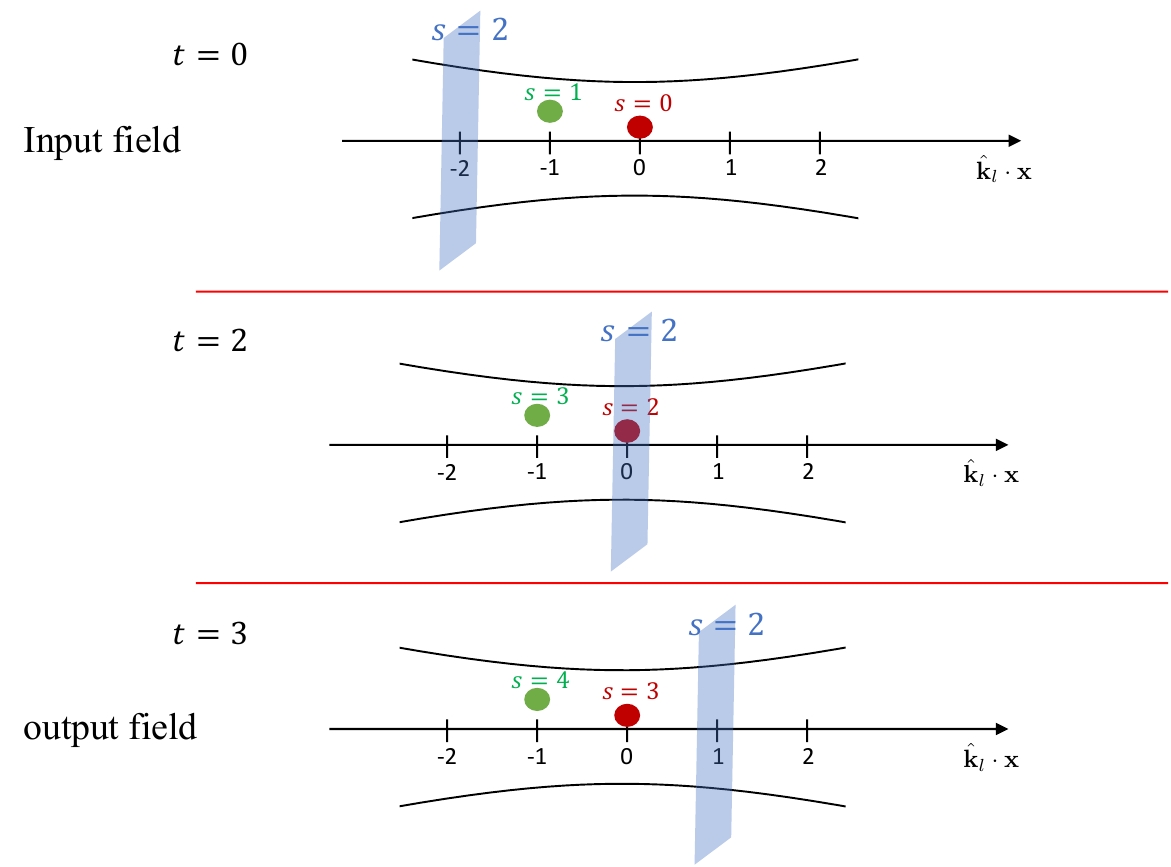}
    \caption{Graphical illustration of the input-output relation. The three snapshots at different times ($t=0,2,3$) show the relationship between the retarded time of a traveling field and the retarded time of molecules fixed in position. For simplicity, we set $c=1$ in this figure. Under the narrow-band approximation, we can define photon field creation and annihilation operators that are localized in retarded time $s=t-\hat{\mathbf{k}}_l\cdot\mathbf{x}$. For example, the field $a_l(s=2)$ is represented by the blue transversal plane across the paraxial spatial mode. It travels at the speed of light in the direction of $\hat{\mathbf{k}}_l$. Two molecules are located at fixed positions $\hat{\mathbf{k}}_l\cdot\mathbf{x}=-1$ and $0$. As time progresses, the retarded time at the positions of the molecules increases. When the retarded time of a field is larger than the retarded time at the position of a molecule (see top panel), the field is called the input field because it has not interacted with the molecule. When the retarded time of a field is smaller than the retarded time at the position of a molecule (see bottom panel), the field is called the output field because the field has finished interacting with the molecule.}
    \label{fig:retarded_time}
\end{figure}

In spectroscopy experiments, the signals are measured in the photon field, and the signal observables can be written in terms of the field operators $a_l^\dagger(s)$ and $a_l(s)$. For example, the photon flux operator $a_l^\dagger(s)a_l(s)$ represents the rate of photons passing through a cross-sectional area in the $l$-th spatial mode (see Fig. (\ref{fig:retarded_time})). The rate is measured at a position and time that correspond to the retarded time $s$. The photon flux has a physical dimension of [1/time]. 
The frequency-dispersed photon count $a^\dagger(\omega)a(\omega)$ represents the density of photons in frequency space, and has a physical dimension of [1/frequency]. Another common observable is the rate of coincidence detection of two photons, one in channel $l_1$ at retarded time $s_1$ and one in channel $l_2$ at retarded time $s_2$. This is described by the operator $a^\dagger_{l_1}(s_1)a^\dagger_{l_2}(s_2)a_{l_2}(s_2)a_{l_1}(s_1)$, having a physical dimension of [1/time$^2$]. These operators can also be convolved with detector response functions to account for the finite time-frequency resolution of photon detectors \cite{Mukamel_Rev_Mod_Phys, del_valle_2012}. 

\section{The input-output relation}
\label{Sec:input_output_relation}
The input-output formalism is based on Hamiltonians of the forms of Eqs. (\ref{Eq:H_int_main}) and (\ref{Eq:H_int_main_many_molecules}), which describe 1-dimensional fields interacting with quantum systems. The input-output relation is one of the main results of the input-output formalism, and it describes how the input and output fields are related.
Traditionally, the input-output relation is derived by converting between the frequency domain and the time domain, and it is sometimes described in the language of quantum stochastic differential equations \cite{Gardiner1985, Gardiner1993, Combes_2017_review, caneva2015quantum}. Here we briefly review the input-output relation. We derive the input-output relation by working only in the (retarded-)time domain and using the language of the more familiar ordinary differential equations. 
\par
The input-output relation is in the Heisenberg picture, and it concerns the $\textbf{time}$-evolution of the $\textbf{retarded time}$ -- dependent field operator $a_l(s)$. 
By a similar logic to Eq. (\ref{Eq:interation_to_Heisenberg_general}), we define the time-evolved retarded time -- dependent field operator as
\begin{equation}
    a_l(s,t) = U^\dagger(t)a_l(s)U(t).
\label{Eq:time-evolved_a}
\end{equation}
Here the retarded time $s$ contains both time and position variables implicitly. For this expression to be physically meaningful, the implicit time variable in $s$ needs to be equal to $t$, hence constraining the position $\mathbf{x}$ of the photon field $a_l(s,t)$ to be on the plane of 
\begin{equation}
    \frac{\hat{\mathbf{k}}_l\cdot\mathbf{x}}{c} = t-s
\label{Eq:retarded_time_as_plane_in_real_space}
\end{equation}
(see Eq. (\ref{Eq:retarded_time_def})). Therefore $a_l(s,t)$ can be thought of as the photon field on the cross section plane $\hat{\mathbf{k}}_l\cdot\mathbf{x}/c=t-s$ at time $t$ (see Fig. \ref{fig:retarded_time}).
\par
To obtain an explicit expression for $a_l(s,t)$, we use the many-molecule Hamiltonian of Eq. (\ref{Eq:H_int_main_many_molecules}) and the Schrodinger equation (Eq. (\ref{Eq:Schrodinger_eqn_1})) to take the partial derivative of $a_l(s,t)$ with respect to $t$ (keeping $s$ constant). The partial derivative is evaluated as
\begin{align}
\begin{split}
    \frac{\partial }{\partial t} a_l(s,t) &= -iU^\dagger(t)\big[a_l(s),H(t)\big]U(t)\\
    &= \sum_{j=1}^N \delta\big(s-s_l(t,\mathbf{x}_j)\big)L_{Hl,j}(t),
\label{Eq:t_derivative_alst}
\end{split}
\end{align}
where $L_{Hl,j}(t)=U^\dagger(t)L_{l,j}(t)U(t)$ is in the Heisenberg picture (see Eq. (\ref{Eq:interation_to_Heisenberg_general})). The delta function is nonzero only when the retarded time $s$ of the field is equal to the retarded time $s_l(t,\mathbf{x}_j)$ at the position of some molecule $j$. If we write $s$ as $t-\hat{\mathbf{k}}_l\cdot\mathbf{x}/c$ (see Eq. (\ref{Eq:retarded_time_as_plane_in_real_space})), then the condition $s=s_l(t,\mathbf{x}_j)$ becomes $\hat{\mathbf{k}}_l\cdot \mathbf{x} = \hat{\mathbf{k}}_l\cdot \mathbf{x}_j$, meaning that the plane of the propagating photon field $a_l(s,t)$ crosses the $j$-th molecule. The condition $s=s_l(t,\mathbf{x}_j)$ is also equivalent to $t=s+\hat{\mathbf{k}}_l\cdot\mathbf{x}_j/c$. Therefore, the field $a_l(s,t)$ changes value (with respect to time propagation) only when $t=s+\hat{\mathbf{k}}_l\cdot\mathbf{x}_j/c$ for some molecule $j$.
\par
We are only interested in the field $a_l(s,t)$ with $s>-\hat{\mathbf{k}}_l\cdot\mathbf{x}_j/c$ for all molecules, so that at $t=0$, $\hat{\mathbf{k}}_l\cdot\mathbf{x}<\hat{\mathbf{k}}_l\cdot\mathbf{x}_j$, and the photon field $a_l(s,t=0)$ is upstream of all molecules in the sample.
Physically, in spectroscopy experiments, the input light is produced at some distance upstream of the matter sample. 
As $t$ increases, the plane of the photon field propagates forward in the $\hat{\mathbf{k}}_l$ direction.
For small enough $t$ such that the plane of the photon field has not interacted with any molecule (corresponding to the condition $s>s_l(t,\mathbf{x}_j)$ for all $j$), the right hand side of Eq. (\ref{Eq:t_derivative_alst}) remains $0$. Therefore in this case, $a_l(s,t)$ is equal to $a_l(s)$, which is independent of $t$ as long as the plane of the photon field has not reached any molecule (see top panel of Fig. (\ref{fig:retarded_time})). We call this the input field
\begin{equation}
    \text{input field} : a_l(s).
\label{Eq:input_field}
\end{equation}
\par
For large enough $t$ such that the plane of the photon field $a(s)$ has propagated past all molecules (corresponding to the condition $s<s_l(t,\mathbf{x}_j)$ for all $j$), the right hand side of Eq. (\ref{Eq:t_derivative_alst}) will remain $0$ for all time afterwards, and $a_l(s,t)$ is again independent of $t$. We call $a_l(s,t)$ in this case the output field, denoted as $a_{l,\text{out}}(s)$. This corresponds to the physical situation where the photon field is detected at some distance downstream of the matter system. From the definition of Eq. (\ref{Eq:time-evolved_a}), we see that $a_{l,\text{out}}(s)$ satisfies the same commutation relations as $a_l(s)$ does, i.e., $[a_{l,\text{out}}(s), a_{l',\text{out}}(s')] = [a^\dagger_{l,\text{out}}(s), a^\dagger_{l',\text{out}}(s')] = 0$ and $[a_{l,\text{out}}(s), a^\dagger_{l',\text{out}}(s')] = \delta(s-s')\delta_{l,l'}$ By integrating over the delta function in Eq. (\ref{Eq:t_derivative_alst}), we see that $a_{l,\text{out}}(s)$ is equal to 
\begin{equation}
    \text{output field}: a_{l,\text{out}}(s)=a_l(s) + \sum_{j=1}^N L_{Hl,j}(s+\frac{\hat{\mathbf{k}}_l\cdot \mathbf{x}_j}{c}).
\label{Eq:input_output_relation}
\end{equation} 
Eq. (\ref{Eq:input_output_relation}) is the input-output relation in the case of $N$ molecules, for the spatial mode $l$. In the case of a single molecule located at position $\mathbf{x} = \mathbf{0}$, the input-output relation simplifies to 
\begin{equation}
    a_{l,\text{out}}(s) = a_l(s) + L_{Hl}(s).
\label{Eq:input_output_relation_single_molecule}
\end{equation}
\par
In classical nonlinear spectroscopy, the output signal's electric field is treated as the sum of the input electric field and the electric field generated by the matter dipole moment \cite{Mukamel_book}. The input-output relation can be thought of as a quantum mechanical version of this statement. It states that the output field is equal to the input field plus the scaled dipole operators $L_{Hl,j}$ in the Heisenberg picture. The scaled dipole operator $L_{Hl,j}$ is evaluated at time $t=s+\hat{\mathbf{k}}_l\cdot\mathbf{x}_j/c$, which is the time when the propagating plane of the photon field $a_l(s)$ reaches the $j$-th molecule. 
We note that, given the Hamiltonian of Eq. (\ref{Eq:H_int_main_many_molecules}), the input-output relation of Eq. (\ref{Eq:input_output_relation}) is an exact result and it holds for arbitrary light-matter interaction strength (i.e., it is non-perturbative).

\par
In Appendix \ref{app:classical_IO}, we derive the input-output relation in the classical setting, using the macroscopic Maxwell's equations. In the classical case, the input-output relation is an equation of complex numbers, not operators. As a consequence, the classical input-output relation cannot properly treat non-classical field states or the entanglement between field and matter.

\section{Examples of analyzing the optical signal using the input-output relation (non-perturbative)}
\label{Sec:spectroscopy_using_IO_examples}
To see how the input-output relation is used in practice, let us consider some examples of output photon field observables in typical QLS setups. In this section, we will consider only the case of a single molecule located at $\mathbf{x}=\mathbf{0}$ for simplicity, and the input-output relation is given by Eq. (\ref{Eq:input_output_relation_single_molecule}).
To consider the case of having many molecules, one simply sum over all molecules using Eq. (\ref{Eq:input_output_relation}). 
We note again that the input-output relation is valid for arbitrary light-matter coupling strength. In Section \ref{Sec:perturbative_expansion_of_output_field}, we shall restrict our attention to the regime of weak light-matter coupling, as this is the case for most molecular spectroscopy experiments, and then expand the input-output relation perturbatively.

\subsection{Photon flux}
One of the most common type of photon field observable is the photon flux \cite{Loudon_2000_book}. In classical spectroscopy experiments, the detection of signals usually takes the form of intensity measurements, and intensity (energy per time in the detection area) is proportion to the photon flux (photon number per time in the detection area). The photon flux signal is defined as $\langle a^\dagger_{l,\text{out}}(s)a_{l,\text{out}}(s)\rangle$, where the expectation value $\langle \cdots \rangle$ is evaluated with respect to the initial state of the matter and the field. We assume the initial state can be factorized into a product of the matter state and the field state. We do not assume the field state is factorizable as a product of states in different spatial modes because the input photons in different spatial modes can be classically correlated or quantum mechanically entangled. 
\par
Using the input-output relation (Eq. (\ref{Eq:input_output_relation})), the output photon flux signal in the $l$-th mode becomes
\begin{equation}
    \langle a^\dagger_{l,\text{out}}(s)a_{l,\text{out}}(s)\rangle = \langle a^\dagger_l(s)a_l(s)\rangle + \langle a^\dagger_l(s)L_{Hl}(s)\rangle + \langle L^\dagger_{Hl}(s)a_l(s)\rangle + \langle L^\dagger_{Hl}(s) L_{Hl}(s)\rangle.
\label{Eq:photon_flux_expansion}
\end{equation}
The first term on the right hand side represents the photon flux of the input light, without any contribution from the matter. It is an expectation value of a purely field operator. Since the initial state is a product state between the matter and the field, the first term becomes an expectation value with respect to just the initial field state. The second and the third terms contain the interference between the input light and the field generated by the matter dipole moment, and these two terms are complex conjugates of each other. These two terms together represent absorption and stimulated emission. The time evolution of the scaled dipole operator $L_{Hl}(s)$ is due to not only the interaction with light in spatial mode $l$, but it can also be due to the interaction with light in other spatial modes. For example, in a pump-probe experiment, the output field is measured in the probe field mode (mode $l$), but the matter interacts with both the pump field mode and the probe field mode.
Note that the Heisenberg-evolved operator $L_{Hl}(s)$ is not purely in the matter degrees of freedom, since the interaction with light mixes the matter and field degrees of freedom. Therefore the expectation value $\langle a^\dagger_l(s)L_{Hl}(s)\rangle$ cannot be factorized as $\langle a^\dagger_l(s)\rangle \langle L_{Hl}(s)\rangle$. This is an important difference to classical light spectroscopy. In classical spectroscopy, the field operator $a_l(s)$ is treated as a complex number instead of an operator, so, for example, the second term factorizes as $a^*_l(s) \langle L_{Hl}(s)\rangle$. Furthermore, in classical spectroscopy, $L_{Hl}(s)$ is purely in the matter degrees of freedom. 
The last term on the right hand side of Eq. (\ref{Eq:photon_flux_expansion}) represents the spontaneous emission, since it contains no direct contribution from the input field.
\par
For arbitrary light-matter coupling strength, the method to evaluate the terms in Eq. (\ref{Eq:photon_flux_expansion}) depends on the specific state of the input light, and in general needs to be considered in a case-by-case basis. To gain insights into the evaluation of these terms, we consider below three different input field states: vacuum state, Fock state, and coherent state. 
\begin{figure}
    \centering
    \includegraphics[scale=0.5]{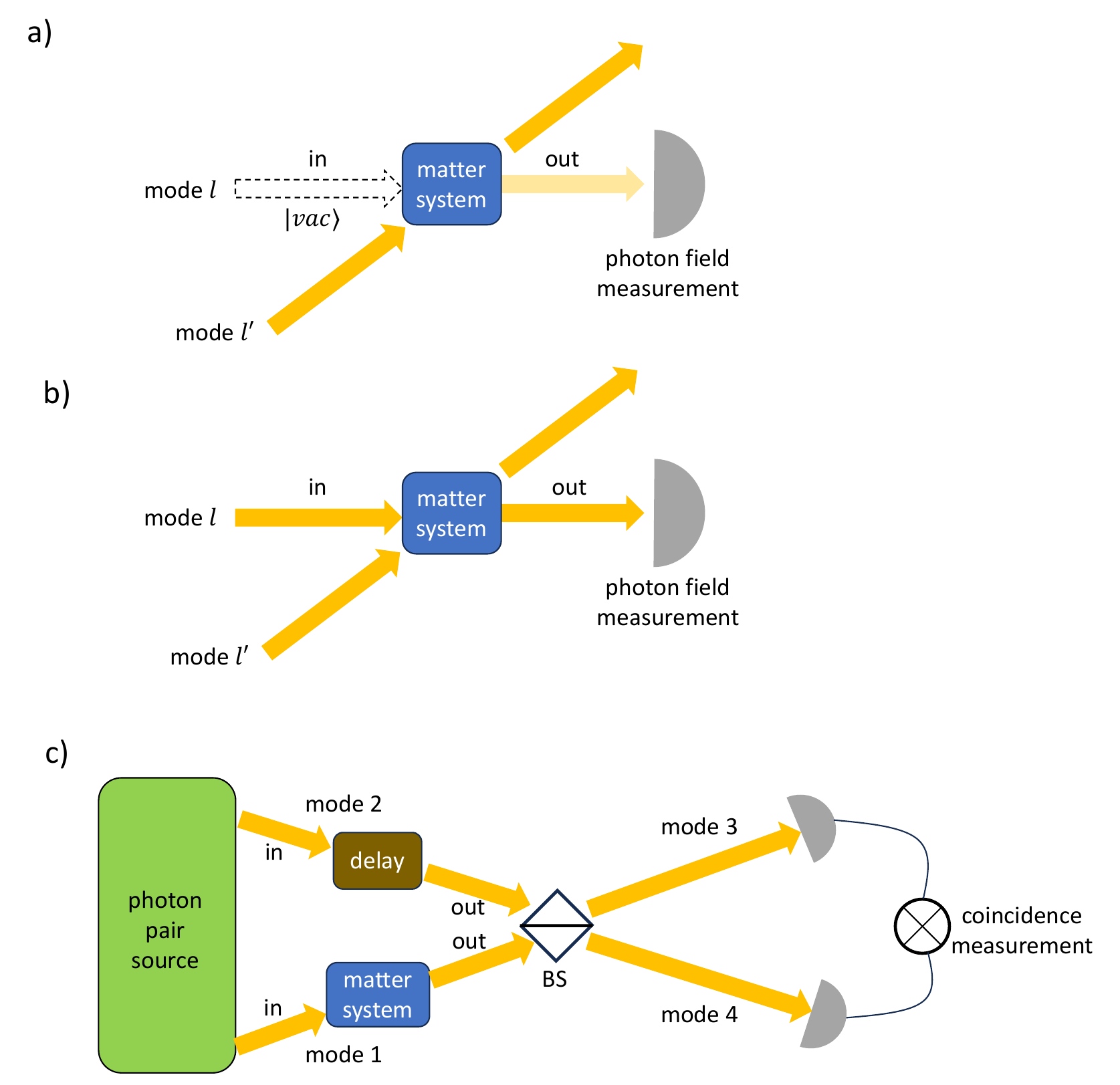}
    \caption{Three common QLS experimental schemes. (a) Measuring the fluorescent light. The matter system interacts with spatial mode $l'$, and the signal is detect in a different spatial mode $l$. The input of mode $l'$ can be any state of light. The input of the detection mode $l$ is the vacuum state. The analysis can be generalized to consider the matter system interacting with more than one non-detection spatial modes. (b) Measuring the transmitted light. Similar to (a), but the input field state of the detection spatial mode $l$ is not a vacuum state. (c) Hong-Ou-Mandel (HOM) scheme. The input field is typically a photon pair with one photon in mode 1 and another in mode 2. The photon in mode 1 interact with a matter system, while the photon in mode 2 is delayed. The two photons interfere with each other in a beamsplitter (BS). The coincidence probability between modes 3 and 4 is measured as a function of the time delay, which provides information about the matter system.}
    \label{fig:measurement_scheme}
\end{figure}
\subsubsection{Vacuum input}
If the spatial mode $l$ is initially in the vacuum state and the matter system is excited by photons in other spatial modes (see Fig. (\ref{fig:measurement_scheme}a)), then the photon flux in the $l$-th spatial mode corresponds to the fluorescence intensity as a function of time. Evaluating the expectation values with respect to the vacuum state, the first three terms in Eq. (\ref{Eq:photon_flux_expansion}) equal to zero. Only the last term (i.e. spontaneous emission) contributes to the output photon flux. To evaluate the expectation value $\langle L^\dagger_{Hl}(s) L_{Hl}(s)\rangle$, one can transform from the Heisenberg picture back into the Schrodinger picture, i.e.,
\begin{align}
\begin{split}
    \langle L^\dagger_{Hl}(s) L_{Hl}(s)\rangle &= \text{Tr}(L^\dagger_l L_l \rho_{\text{tot}}(s)) \\
    &= \text{Tr}_{\text{sys}}(L^\dagger_l L_l \rho_{\text{sys}}(s)),
\label{Eq:photon_flux_vacuum}
\end{split}
\end{align}
where $\rho_{\text{sys}}=\text{Tr}_{\text{field}}(\rho_{\text{tot}})$ is the reduced density matrix for the system. On the right hand side of the first equality, the Schrodinger picture operator $L^\dagger_l L_l$ only acts on the system degrees of freedom. Therefore in the second equality, we can trace out the field degrees of freedom from $\rho_{\text{sys+field}}$. Note that the photon flux expectation value is now re-expressed as an expectation value in the system degrees of freedom. For an arbitrary initial field state, there is no general method to evaluate $\rho_{\text{sys}}(t)$ in the non-perturbative regime, but there are different master equations to treat the effects of different input states \cite{wiseman_milburn_2009_book,Gardiner1985,Baragiola_2012}. In the case that the matter system reaches a steady state due to a stationary input from a spatial mode other than $l$, then the photon flux expectation value is obtained by substituting the system steady state $\rho_{\text{steady}}$ into Eq. (\ref{Eq:photon_flux_vacuum}). In Appendix \ref{app:spontaneous_emission_Fermi_rate}, we show that Eq. (\ref{Eq:photon_flux_vacuum}) is consistent with the the spontaneous emission rate given by Fermi's golden rule in a simple example.

\subsubsection{$m$-photon Fock state input}
Fig. (\ref{fig:measurement_scheme}b) illustrates the situation when the input field state in mode $l$ is not a vacuum state. As an example, we consider an $m$-photon Fock state as the input in the $l$-th mode.
An $m$-photon Fock state is defined as
\begin{equation}
    |m_\xi\rangle_l = \frac{1}{\sqrt{m!}} \Big(\int ds\, \xi(s)a_l^\dagger(s) \Big)^m|\text{vac}\rangle,
\label{Eq:m_photon_Fock_state_1}
\end{equation}
where $\xi(s)$ is the temporal profile of the pulse, and $|\text{vac}\rangle$ means the vacuum state. $\xi$ is normalized such that $\int ds\,|\xi(s)|^2 = 1$. The last term in Eq. (\ref{Eq:photon_flux_expansion}) is evaluated in a similar way as in Eq. (\ref{Eq:photon_flux_vacuum}). The first term in Eq. (\ref{Eq:photon_flux_expansion}) is evaluated with respect to the initial field state of Eq. (\ref{Eq:m_photon_Fock_state_1}), and it is equal to $m|\xi(s)|^2$. The second and the third term in Eq. (\ref{Eq:photon_flux_expansion}) require a more careful treatment. To do so, we first write the combined initial state as $\rho_{\text{tot}}(0) = \rho_{\text{sys}}(0)\otimes \rho_{\text{field},\neq l}\otimes |m_\xi\rangle_l\langle m_\xi|$, where $\rho_{\text{field},\neq l}$ is the field state of all modes other than $l$. Then the second term (similarly for the third term) becomes
\begin{align}
\begin{split}
    \langle a^\dagger_l(s)L_{Hl}(s)\rangle &= \text{Tr}\Big(L_{Hl}(s)\big(\rho_{\text{sys}}(0)\otimes \rho_{\text{field},\neq l}\otimes |m_\xi\rangle_l\langle m_\xi|\big)a^\dagger_l(s) \Big) \\
    &= \sqrt{m}\xi^*(s)\text{Tr}\Big(U^\dagger(s) L_l U(s)\big(\rho_{\text{sys}}(0)\otimes \rho_{\text{field},\neq l}\otimes |m_\xi\rangle_l\langle (m-1)_\xi|\big) \Big)\\
    &= \sqrt{m}\xi^*(s)\text{Tr}\Big(L_l \mathcal{G}(s)\big(\rho_{\text{sys}}(0)\otimes \rho_{\text{field},\neq l}\otimes |m_\xi\rangle_l\langle (m-1)_\xi|\big) \Big)\\
    &= \sqrt{m}\xi^*(s)\text{Tr}_{\text{sys}}\big(L_l \rho_{m,m-1}(s)\big).
\label{Eq:photon_flux_Fock_2nd_term}
\end{split}
\end{align}
In the first equality, we have used the invariance of trace under cyclic permutation. In the second equality, we have used the photon annihilation property $a_l(s)|m_\xi\rangle_l = \sqrt{m}\xi(s)|(m-1)_\xi\rangle_l$ and that $L_{Hl}(s)=U^\dagger(s)L_l(s)U(s)$. In the third equality, we define the time evolution superoperator $\mathcal{G}(s)$, whose action on an operator $X$ is given by $\mathcal{G}(s)X = U(s)XU^\dagger(s)$. In the final equality, we define the auxiliary system density matrix $\rho_{m,n}(s)$ as $\text{Tr}_{\text{field}}(\mathcal{G}(s)(\rho_{\text{sys}}(0)\otimes \rho_{\text{field},\neq l}\otimes |m_\xi\rangle_l\langle n_\xi|))$, i.e., the reduced system density matrix for the time-evolved non-physical state $\rho_{\text{sys}}(0)\otimes \rho_{\text{field},\neq l}\otimes |m_\xi\rangle_l\langle n_\xi|$. Note that by definition, the physical reduced system state $\rho_{\text{sys}}(t) = \text{Tr}_{\text{field}}\rho_{\text{tot}}(t)$ is $\rho_{m,m}(t)$. For arbitrary $\rho_{\text{field},\neq l}$, there is no general method to evaluate Eq. (\ref{Eq:photon_flux_Fock_2nd_term}) in the non-perturbative regime. 
\par
In the simplest case, we take $\rho_{\text{field},\neq l}$ to be the vacuum state. Then the equations of motion of the auxiliary density matrices $\rho_{m,n}$ are described by the hierarchy of Fock state master equations (see \cite{Baragiola_2012, Ko_2022} or Sec. \ref{sec:Fock_state_normal_ordered_expansion}). 
We note that this method of calculating the photon flux was first derived in \cite{Baragiola_2012} using the mathematics of quantum stochastic differentials, and we have re-formulated the derivation in terms of ordinary calculus in \cite{Ko_2022}. 

\subsubsection{Coherent state input}
A coherent state with a coherent amplitude $\alpha(s)$ in mode $l$ is defined as
\begin{equation}
    |\alpha\rangle_l = \exp \big(\int ds\, \alpha(s)a_l^\dagger(s) - \alpha^*(s)a_l(s) \big) |\text{vac}\rangle,
\label{Eq:coherent_state_def_IO}
\end{equation}
and it has the property $a_l(s)|\alpha\rangle = \alpha(s)|\alpha\rangle$.
If this is used as the input in spatial mode $l$, then Eq. (\ref{Eq:photon_flux_expansion}) becomes
\begin{equation}
    \langle a^\dagger_{l,\text{out}}(s)a_{l,\text{out}}(s)\rangle = |\alpha(s)|^2 + \alpha^*(s) \langle L_{Hl}(s)\rangle + \alpha(s) \langle L^\dagger_{Hl}(s)\rangle + \langle L^\dagger_{Hl}(s) L_{Hl}(s)\rangle,
\label{Eq:photon_flux_coherent}
\end{equation}
where the $a_l(s)$ and $a^\dagger_l(s)$ in Eq. (\ref{Eq:photon_flux_expansion}) are replaced with $\alpha(s)$ and $\alpha^*(s)$. The expectation values $\langle L_{Hl}(s)\rangle$, $\langle L^\dagger_{Hl}(s)\rangle$, and $\langle L^\dagger_{Hl}(s) L_{Hl}(s)\rangle$ in Eq. (\ref{Eq:photon_flux_coherent}) are evaluated in a similar way as in Eq. (\ref{Eq:photon_flux_vacuum}). 
\par
The system density matrix in the Schrodinger picture $\rho_{\text{sys}}(t)$ is needed to evaluate the expectation values, but again, there is no general method to evaluate $\rho_{\text{sys}}(t)$ for an arbitrary initial field state $\rho_{\text{field},\neq l}$ in modes other than $l$. If $\rho_{\text{field},\neq l}$ is the vacuum state, then $\rho_{\text{sys}}$ can be obtained by solving the coherent state master equation \cite{gardiner_zoller_quantum_noise, wiseman_milburn_2009_book, Ko_2022}
\begin{equation}
    \frac{d\rho_{\text{sys}}}{dt} = [-iH_{\text{sys}}-\alpha(t)L_l^\dagger + \alpha^*(t)L_l, \rho_{\text{sys}}] + \sum_{l'} \Big( L_{l'}\rho_{\text{sys}} L_{l'} - \frac{1}{2}L_{l'}^\dagger L_{l'} \rho_{\text{sys}} - \frac{1}{2}\rho_{\text{sys}} L_{l'}^\dagger L_{l'}\Big) .
\end{equation}

\subsection{Intra-mode second order photon coherence function}
The second order photon coherence function $g^{(2)}$ has been increasingly recognized as a useful observable that can reveal information about the matter system, especially when its value lies in the non-classical regime \cite{lupton2021review, munoz_g2_2020, Holdaway_2018, Kalashnikov_2017_HOM, asban2021distinguishability, asban2022nonlinear}. We consider the intra-mode correlation in this section, and discuss an example of the inter-mode correlation in Sec. \ref{sec:HOM}.
\par
The intra-mode, un-normalized, second-order photon coherence function of mode $l$ output is defined as
\begin{equation}
    G_{l,l}^{(2)}(s_1, s_2) = \big\langle a_{l,\text{out}}^\dagger(s_1) a_{l,\text{out}}^\dagger(s_2) a_{l,\text{out}}(s_2) a_{l,\text{out}}(s_1) \big\rangle.
\label{Eq:G_2_1}
\end{equation}
Due to the commutation relations of $a_{l,\text{out}}(s)$ (see texts above Eq. (\ref{Eq:input_output_relation})), $G^{(2)}_{l,l}(s_1,s_2) = G^{(2)}_{l,l}(s_2,s_1)$.
Therefore, without loss of generality, we shall impose the time-ordering $s_2>s_1$ in Eq. (\ref{Eq:G_2_1}) \cite{Holdaway_2018}. Physically, $G^{(2)}(s_1,s_2)dt^2$ is the joint probability of observing a photon at $(s_1,s_1+dt)$ and observing another photon at $(s_2, s_2+dt)$, where $dt$ is the infinitesimal time increment.
The normalized second order coherence function, $g_{l,l}^{(2)}(s_1,s_2)$, is related to $G_{l,l}^{(2)}(s_1,s_2)$ by $g_{l,l}^{(2)}(s_1,s_2)=G_{l,l}^{(2)}(s_1,s_2)/(\langle a_{l}^\dagger(s_1)a_{l}(s_1)\rangle \langle a_{l}^\dagger(s_2)a_{l}(s_2)\rangle)$, where we have dropped the subscript ``out" for generality. 
Using the input-output relation (Eq. (\ref{Eq:input_output_relation_single_molecule})), Eq. (\ref{Eq:G_2_1}) becomes
\begin{align}
\begin{split}
    G^{(2)}_{l,l}(s_1,s_2) = \Big\langle & \big(a^\dagger_l(s_1)+L^\dagger_{Hl}(s_1)\big)\big(a^\dagger_l(s_2)+L^\dagger_{Hl}(s_2)\big) \\
    & \big(a_l(s_2)+L_{Hl}(s_2)\big)\big(a_l(s_1)+L_{Hl}(s_1)\big)\Big\rangle.
\label{Eq:G_2_using_IO_relation}
\end{split}
\end{align}
For arbitrary light-matter coupling strength, this expression needs to be evaluated in a case-by-case basis, depending on the state of the input light. 
\par
If one measures the photon correlation $G^{(2)}_{l,l}$ of the fluorescent photon, then the input state in mode $l$ is the vacuum state (see Fig. (\ref{fig:measurement_scheme}a)). It has been shown that the fluorescence $g^{(2)}$ of a molecular aggregate can be used as an indicator for the quantum coherence between individual chromophores \cite{munoz_g2_2020, Holdaway_2018}.
When the expectation value in Eq. (\ref{Eq:G_2_using_IO_relation}) is evaluated with respect to the vacuum state in mode $l$, terms involving $a_l(s)$ or $a_l^\dagger(s)$ in Eq. (\ref{Eq:G_2_using_IO_relation}) will be zero, so $G^{(2)}_{l,l}(s_1,s_2)$ reduces to
\begin{align}
\begin{split}
    G_{l,l}^{(2)}(s_1, s_2) &= \Big\langle L_{Hl}^\dagger(s_1)L_{Hl}^\dagger(s_2)L_{Hl}(s_2)L_{Hl}(s_1)\Big\rangle\\
    &=\text{Tr}\Big(L_{Hl}^\dagger(s_2)L_{Hl}(s_2)L_{Hl}(s_1) \rho_{\text{tot}}(0)L_{Hl}^\dagger(s_1) \Big)\\
    &=\text{Tr}\Big(L_{l}^\dagger L_{l} \mathcal{G}(s_2-s_1)\big(L_{l}\, \rho_{\text{tot}}(s_1)L_{l}^\dagger\big) \Big).
\label{Eq:G_2_2}
\end{split}
\end{align}
We note that the trace is performed in the matter and field degrees of freedom, which include photon mode $l$ that initializes in the vacuum state and all other photon modes that excite the matter system.
To obtain the third line from the second line, one expands the Heisenberg picture operators $A_H(s)$ as $e^{iH_{\text{sys+field}}t}Ae^{-iH_{\text{sys+field}}t}$ and use the invariance of trace under cyclic permutation. The final line in Eq. (\ref{Eq:G_2_2}) has an intuitive physical interpretation: As a photon is observed at time $s_1$, the combined state $\rho_{\text{tot}}(s_1)$ undergoes a de-excitation jump to the (un-normalized) state $L_l\rho_{\text{tot}}(s_1)L^\dagger_l$. 
After evolving this state from $s_1$ to $s_2$ (described by $\mathcal{G}(s_2-s_1)$), another photon is observed, corresponding to another de-excitation jump. The joint probability of observing two photons is obtained by taking the trace at the end.
\par
The fluorescent $G^{(2)}_{l,l}(s_1,s_2)$ in Eq. (\ref{Eq:G_2_2}) can be further simplified if the matter system is driven by a stationary photon source in mode $l'$ (i.e., a mode other than $l$), and if the combined system+field state reaches a steady state. In this case, $G^{(2)}_{l,l}(s_1,s_2)$ only depends on the time difference $s=s_2-s_1$, so we write the second order coherence function as $G^{(2)}_{l,l}(s)$. The total state $\rho_{\text{tot}}(s_1)$ is replaced with $\rho_{\text{steady}}$, the steady state in the combined system+field degrees of freedom. Eq. (\ref{Eq:G_2_2}) now becomes
\begin{equation}
    G^{(2)}_{l,l}(s) = \text{Tr}\Big(L_{l}^\dagger L_{l} \mathcal{G}(s)\big(L_{l}\, \rho_{\text{steady}}L_{l}^\dagger\big) \Big).
\label{Eq:G2_steady}
\end{equation} 
From this equation, we see that the steady state $G^{(2)}_{l,l}(s)$ contains information about the transient dynamics of the perturbed steady state $L_l\rho_{\text{steady}}L_{l}^\dagger$.

\subsection{Hong-Ou-Mandel (HOM) interferometry}
\label{sec:HOM}
As an example of the inter-mode second order coherence function, we consider the Hong-Ou-Mandel (HOM) interferometry scheme in Fig. (\ref{fig:measurement_scheme}c). A photon pair with one photon in spatial mode $1$ and the other photon in spatial mode $2$ is used as the input. The photon pair state 
\begin{equation}
    |\Psi\rangle = \int ds_1 ds_2 \,f(s_1,s_2) a^\dagger_1(s_1)a^\dagger_2(s_2)|\text{vac}\rangle
\label{Eq:biphoton_wavefunction}
\end{equation}
is specified by the biphoton wavefunction $f(s_1,s_2)$. Normalization of $|\Psi\rangle$ requires that $f(s_1,s_2)$ be normalized as $\int ds_1\,ds_2 |f(s_1,s_2)|^2 = 1$. In the HOM scheme, one photon (photon in mode $1$ in Fig. (\ref{fig:measurement_scheme}c)) in the photon pair interacts with a matter system, while the other photon (photon in mode $2$ in Fig. (\ref{fig:measurement_scheme}c)) propagates freely with a time delay $\tau$. The relative time delay $\tau$ between the two photons is varied in the HOM experiment. The output photons in modes 1 and 2 pass through a 50:50 beamsplitter and transform into photons in modes 3 and 4. The beamsplitter transformation is given by
\begin{equation}
    \begin{pmatrix}
        a_3(s) \\
        a_4(s)
    \end{pmatrix}
    = \frac{1}{\sqrt{2}}
    \begin{pmatrix}
        1 & i \\
        i & 1
    \end{pmatrix}
    \begin{pmatrix}
        a_{1,\text{out}}(s) \\
        a_{2,\text{out}}(s)
    \end{pmatrix}.
\label{Eq:beamsplitter_transformation}
\end{equation}
Given a pair of input photon, the probability $P_{34}(\tau)$ of observing a coincidence count in modes 3 and 4 are measured as a function of the time delay $\tau$.
\par
The coincidence probability is given by the expectation value
\begin{equation}
    P_{34}(\tau) = \int ds\, ds' \big\langle a^\dagger_3(s)a^\dagger_4(s')a_4(s')a_3(s)\big\rangle.
\end{equation}
Using the beamsplitter transformation (Eq. (\ref{Eq:beamsplitter_transformation})) and using the fact that mode 1 and mode 2 each contains at most one photon, $P_{34}(\tau)$ becomes
\begin{align}
\begin{split}
    P_{34}(\tau) &= \frac{1}{4} \int ds\, ds' \Big\langle \big(a_{1,\text{out}}^\dagger(s)a_{2,\text{out}}^\dagger(s')-a_{1,\text{out}}^\dagger(s')a_{2,\text{out}}^\dagger(s)\big)\\
    &\qquad\qquad\qquad\quad\big( a_{2,\text{out}}(s')a_{1,\text{out}}(s)-a_{2,\text{out}}(s)a_{1,\text{out}}(s')\big) \Big\rangle \\
    &=\frac{1}{2}\int ds\,ds' \Big\langle a_{1,\text{out}}^\dagger(s)a_{2,\text{out}}^\dagger(s')a_{2,\text{out}}(s')a_{1,\text{out}}(s)\Big\rangle \\
    &\qquad\qquad\qquad-\Big\langle a_{1,\text{out}}^\dagger(s)a_{2,\text{out}}^\dagger(s')a_{2,\text{out}}(s)a_{1,\text{out}}(s')\Big\rangle.
\end{split}
\end{align}
Since the output $a_{2,\text{out}}(s)$ has a time delay $\tau$ relative to the input $a_2(s)$, it is related to the input field by $a_{2,\text{out}}(s)=a_2(s-\tau)$. Combining this with the input-output relation for mode 1 (i.e., $a_{1,\text{out}}(s)=a_1(s)+L_{H1}(s)$), we have
\begin{align}
\begin{split}
    P_{34}(\tau) = \frac{1}{2}\int ds\,ds' &\Big\langle \big(a^\dagger_1(s)+L^\dagger_{H1}(s)\big)a^\dagger_2(s'-\tau) a_2(s'-\tau)\big( a_1(s)+L_{H1}(s) \big)\Big\rangle\\
    &-\Big\langle \big(a^\dagger_1(s)+L^\dagger_{H1}(s)\big)a^\dagger_2(s'-\tau) a_2(s-\tau)\big( a_1(s')+L_{H1}(s') \big)\Big\rangle.
\label{Eq:HOM_IO}
\end{split}
\end{align} 
\par
In the absence of light-matter coupling, $L_{H1}$ is equal to $0$. Then evaluating Eq. (\ref{Eq:HOM_IO}) using the photon pair state of Eq. (\ref{Eq:biphoton_wavefunction}), we see that
\begin{align}
\begin{split}
    P_{34}(\tau) &= \frac{1}{2} \int ds \, ds' f^*(s,s'-\tau)f(s,s'-\tau)-f^*(s,s'-\tau)f(s',s-\tau)\\
    &=\frac{1}{2}\Big(1-\int ds \, ds'f^*(s,s')f(s'+\tau,s-\tau)\Big).
\label{Eq:P34_no_matter}
\end{split}
\end{align}
If the biphoton wavefunction is symmetric (i.e., $f(s_1, s_2)=f(s_2,s_1)$), then $P_{34}(0)$ is zero and $P_{34}(\tau)$ is a symmetric function of $\tau$. To see that the integral in the last line is symmetric in $\tau$, one can switch the time arguments in $f(s_1,s_2)$ (i.e., $f(s_1,s_2)\rightarrow f(s_2,s_1)$), and then switch the variable names $(s,s')\rightarrow(s',s)$. If the photon in mode 1 interacts with a matter system, $P_{34}(0)$ will no longer be $0$, and $P_{34}(\tau)$ will no longer be symmetric. The behavior of $P_{34}(\tau)$ can be used to determine the properties of the matter system, such as the dephasing time \cite{Kalashnikov_2017_HOM}.

\section{Perturbative expansion of the optical signal}
\label{Sec:perturbative_expansion_of_output_field}
Typical molecular spectroscopy experiments operate in the weak coupling perturbative regime, meaning that the energy scale of the light-matter interaction strength is small compared to other energy scales relevant to the dynamics. The perturbative approach for quantum light spectroscopy provides a unified method to treat general photon input states under the weak coupling regime.
To analyze the output optical signal perturbatively, we perform a perturbative expansion on the input-output relation (Eqs. (\ref{Eq:input_output_relation}) and (\ref{Eq:input_output_relation_single_molecule})) by expanding the Heisenberg picture operator $L_{Hl}(s)$ in terms of purely system and purely field operators.
\par
The conventional perturbative approach for quantum light spectroscopy works in the interaction picture, where the system+field state $\rho_{\text{tot}}(t)$ is expanded perturbatively, and then the expectation values are evaluated with respect to the time-evolved system+field state \cite{Mukamel_Rev_Mod_Phys, Schlawin_2017_tutorial}. One can alternatively view the conventional perturbing-the-state approach in the interaction picture as a perturbing-the-observable approach in the Heisenberg picture, a point of view that is more closely related to our input-output approach. In this section, we will discuss these two perspectives of the conventional perturbative approach, and then present our input-output approach that perturbs the input-output relation in the Heisenberg picture.
We argue that the input-output formulation is more natural and intuitive for analyzing the optical signal. It has led us to discover an equivalence between a class of quantum light spectroscopy experiments using entangled biphotons and a class of classical spectroscopy experiments \cite{Ko_2023}.
We will see that the input-output approach also avoids an issue regarding a subtle integration bound in the conventional approach, which has not been addressed in the literature.

\par
We will use the notational convention described in Sec. \ref{sec:interaction_pic_Hamiltonian}. To remind the readers, $H(t)$ is the time-dependent interaction picture Hamiltonian, and $U(t)$ is the time evolution operator generated by $H(t)$. An operator without subscript $H$ (e.g., $A(t)$ or $H(t)$) is understood to be in the interaction picture. An operator with subscript $H$ (e.g., $A_H(t)$ or $H_H(t)$) is understood to be in the Heisenberg picture. A Heisenberg picture operator $A_H(t)$ is related to its corresponding interaction picture operator $A(t)$ by $A_H(t) = U^\dagger(t)A(t)U(t)$.

\subsection{Perturbative expansion of Heisenberg-evolved operators}
\label{Sec:perturbative_expansion_of_Heisenberg_operator}

As a preliminary to the following discussion, we first show how to perturbatively expand a Heisenberg picture operator $A_H(t)$ in terms of interaction picture operators. 
We show in Appendix \ref{app:Heisenberg_perturbation} that a time-ordered expansion of $A_H(t)$ from $A(t)$ produces the result
\begin{align}
\begin{split}
    A_H(t)=& A(t)-i\int^t_{0}dt_1\,[A(t),H_H(t_1)] \\
    &+(-i)^2\int^t_{0}dt_2\int^{t_2}_{0}dt_1\,[[A(t),H_H(t_1)],H_H(t_2)]\\
    &+(-i)^3\int^t_{0}dt_3\int^{t_3}_{0}dt_2\int^{t_2}_{0}dt_1\,[[[A(t),H_H(t_1)],H_H(t_2)],H_H(t_3)]+\cdots.
\label{A_H_expansion_1_time_ordered}
\end{split}
\end{align}
The commutators with the Hamiltonian are applied in a time-ordered manner, where the Hamiltonians with smaller time arguments are applied first. However, this is not a useful expansion for analyzing the optical signal because the Hamiltonians in the expansion are in the Heisenberg picture, not in the interaction picture. In the context of quantum light spectroscopy, the Heisenberg-evolved $H_H(t)$ is a complicated object that mixes the matter and the field degrees of freedom. We would like to have an expansion in terms of interaction picture operators, which can be decomposed into purely system and purely field operators. We note that a different time-ordered expansion of a Heisenberg-evolved operator $A_H(t)$ from $A_H(0)$ has been derived in Appendix 5B of Ref. \cite{Mukamel_book} using the Liouville space representation, in the context of classical spectroscopy. 
\par
The perturbative expansion that we will use to expand a Heisenberg-evolved operator $A_H(t)$ is the following:
\begin{align}
\begin{split}
    A_H(t)= A(t) &-i\int^t_0 dt_1\,[A(t),H(t_1)] \\
    &+ (-i)^2\int^t_0 dt_2 \int^{t_2}_0 dt_1\,[[A(t),H(t_2)],H(t_1)] \\
    &+ (-i)^3\int^t_0 dt_3 \int^{t_3}_0 dt_2 \int^{t_2}_0 dt_1\,[[[A(t),H(t_3)],H(t_2)],H(t_1)] +\cdots ,
\label{Eq:A_H_perturbative_expansion}
\end{split}
\end{align}
Details of the derivation is shown in Appendix \ref{app:Heisenberg_perturbation}.
Notice that this expansion is anti-time-ordered in the sense that Hamiltonians with larger time arguments are applied first. Furthermore, in contrast to the expansion in Eq. (\ref{A_H_expansion_1_time_ordered}), the Hamiltonians in this expansion are in the interaction picture, allowing us to decompose each term into purely system and purely field operators (see Eqs. (\ref{Eq:H_int_main}) and (\ref{Eq:H_int_main_many_molecules})).

\subsection{Three different perturbative approaches to analyze optical signals}
\label{Sec:conventional_perturbative_approach}
\subsubsection{Approach 1: perturbing the state in the interaction picture (conventional method)}
\label{sec:conventional_pert_interaction_pic}
In the conventional perturbative approach, one perturbs the combined system+field density matrix $\rho_{\text{tot}}(t)$ in the interaction picture. The perturbative series is written as
\begin{align}
\begin{split}
    \rho_{\text{tot}}(t) = & \rho(0) -i\int^t_{0}dt_1\,[H(t_1),\rho_{\text{tot}}(0)]\\
    & + (-i)^2\int^t_{0}dt_2 \int^{t_2}_{0} dt_1\, [H(t_2),[H(t_1),\rho_{\text{tot}}(0)]] \\
    & + (-i)^3 \int^t_{0} dt_3\int^{t_3}_{0}dt_2\int^{t_2}_{0}dt_1\, [H(t_3),[H(t_2),[H(t_1),\rho_{\text{tot}}(0)]]] + \cdots.
\label{Eq:rho_perturbative_series}
\end{split}
\end{align}
As discussed in Sec. \ref{sec:interaction_pic_Hamiltonian} and \ref{sec:photon_field_observables}, in the interaction picture, photon field observables are expressed as functions of the retarded time $s$. The expectation value of a photon field observable $A(s)$ is given by $\text{Tr}( A(s)\rho_{\text{tot}}(t))$.
Substituting Eq. (\ref{Eq:rho_perturbative_series}) into this expectation value, we obtain a perturbative expansion of the optical signal
\begin{align}
\begin{split}
    \langle A(s) \rangle
    = & \,\text{Tr}\big(A(s)\rho_{\text{tot}}(0)\big) -i\int^t_{0}dt_1\,\text{Tr}\big(A(s)[H(t_1),\rho_{\text{tot}}(0)]\big)\\
    & + (-i)^2\int^t_{0}dt_2 \int^{t_2}_{0} dt_1\, \text{Tr}\big(A(s)[H(t_2),[H(t_1),\rho_{\text{tot}}(0)]]\big) \\
    & + (-i)^3 \int^t_{0} dt_3\int^{t_3}_{0}dt_2\int^{t_2}_{0}dt_1\, \text{Tr}\big(A(s)[H(t_3),[H(t_2),[H(t_1),\rho_{\text{tot}}(0)]]]\big) + \cdots.
\label{Eq:A_expectation_1}
\end{split}
\end{align}
\par
Since the optical signal $A(s)$ at retarded time $s$ is measured after the photon field has traveled past the matter system, we require that the time $t$ in Eq. (\ref{Eq:A_expectation_1}) be large enough such that the plane of the photon field $a(s)$ has propagated past all molecules. Mathematically, this condition is expressed as $s_l(t,\mathbf{x}_j)=t-\hat{\mathbf{k}}_l\cdot\mathbf{x}_j/c > s$ for all molecules $j$ (see Fig. (\ref{fig:retarded_time}) for an illustration and Sec. \ref{Sec:input_output_relation} for more detailed discussion on the retarded time). If the field observable involves more than one retarded time variable or more than one field mode (e.g., the second order coherence function $a_l^\dagger(s_1)a_{l'}^\dagger(s_2)a_{l'}(s_2)a_l(s_1)$), then one needs to make sure that $t$ is large enough such that $s_l(t,\mathbf{x}_j)$ is greater than all retarded time variables $s_i$, for all spatial modes $l_i$. For convenience, $t$ can simply be set to $\infty$. In the literature \cite{Mukamel_Rev_Mod_Phys, Schlawin_2017_tutorial}, the distinction between the interaction time $t$ and the retarded time $s$ is often overlooked. This has resulted in inconsistencies in the choice of the time $t$ in the literature. Sometimes, $t$ is set to be the retarded time $s$, while other times, $t$ is set to be $\infty$. In the input-output approach, the choice of the integration time is automatically taken care of by the input-output relation. We will discuss this issue in more detail using an example in Sec. \ref{Sec:comparing_IO_to_conventional}.

\subsubsection{Approach 2: perturbing the observable in the Heisenberg picture}
\label{Sec:conventional_perturbation_Heisenberg}
The expectation value $\text{Tr}(A(s)\rho_{\text{tot}}(t)) = \text{Tr}(A(s)U(t)\rho_{\text{tot}}(0)U^\dagger(t))$ can be expressed in the Heisenberg picture by cyclically permuting the $U^\dagger(t)$. The expectation value then becomes $\text{Tr}(U^\dagger(t)A(s)U(t)\rho_{\text{tot}}(0))$, where $U^\dagger(t)A(s)U(t)$ is the time-evolved field operator in the Heisenberg picture. In contrast to the first approach where we expand the state $\rho_{\text{tot}}(t) = U(t)\rho_{\text{tot}}(t)U^\dagger(t)$ in the interaction picture, we can alternatively expand the Heisenberg picture operator $U^\dagger(t)A(s)U(t)$ using Eq. (\ref{Eq:A_H_perturbative_expansion}). The point of view taken by this approach is more closely related to our input-output method, i.e., the third approach. 
\par
Noting that the retarded time $s$ is implicitly a function of time $t$ and position $\mathbf{x}$, we can apply Eq. (\ref{Eq:A_H_perturbative_expansion}) to expand $U^\dagger(t)A(s)U(t)$ as
\begin{align}
\begin{split}
    U^\dagger(t)A(s)U(t)= & A(s) -i\int^t_0 dt_1\,[A(s),H(t_1)] \\
    &+ (-i)^2\int^t_0 dt_2 \int^{t_2}_0 dt_1\,[[A(s),H(t_2)],H(t_1)] \\
    &+ (-i)^3\int^t_0 dt_3 \int^{t_3}_0 dt_2 \int^{t_2}_0 dt_1\,[[[A(s),H(t_3)],H(t_2)],H(t_1)] +\cdots .
\label{Eq:A_s_t_perturbative_expansion}
\end{split}
\end{align}
Taking the trace of Eq. (\ref{Eq:A_s_t_perturbative_expansion}) with respect to the initial state $\rho_{\text{tot}}(0)$, we have
\begin{align}
\begin{split}
    \langle A(s)\rangle = & \text{Tr}\big( A(s)\rho_{\text{tot}}(0)\big) -i\int^t_{0}dt_1\,\text{Tr}\big([A(s), H(t_1)]\rho_{\text{tot}}(0)\big)\\
    & + (-i)^2\int^t_{0}dt_2 \int^{t_2}_{0} dt_1\, \text{Tr}\big([[A(s), H(t_2)],H(t_1)]\rho_{\text{tot}}(0)\big) \\
    & + (-i)^3 \int^t_{0} dt_3\int^{t_3}_{0}dt_2\int^{t_2}_{0}dt_1\, \text{Tr}\big([[[A(s), H(t_3)],H(t_2)],H(t_1)]\rho_{\text{tot}}(0)\big)  + \cdots.
\label{Eq:A_expectation_perturbation}
\end{split}
\end{align}
By applying the identity $\text{Tr}([A,B]C)=\text{Tr}(A[B,C])$ repeatedly, we can see that Eq. (\ref{Eq:A_expectation_perturbation}) is equal to Eq. (\ref{Eq:A_expectation_1}), confirming the equivalence to the conventional perturbing-the-state approach. 

\subsubsection{Approach 3: perturbative expansion of the input-output relation}
\label{Sec:IO_perturbation}
A different way to express the Heisenberg-evolved field operator $U^\dagger(t)A(s)U(t)$ is to replace every photon creation or annihilation operators in $A(s)$ with the corresponding output field operator. For example, consider $A(s) = a_l^\dagger(s)a_l(s)$. In the Heisenberg picture, $U^\dagger(t)a^\dagger_l(s)a_l(s)U(t) = \big(U^\dagger(t)a^\dagger_l(s)U(t)\big)\big(U^\dagger(t)a_l(s)U(t)\big)$. We identify $U^\dagger(t)a_l(s)U(t)$ as the output field $a_{l,\text{out}}(s)$ because $t$ is assumed to be large enough such that the field at retarded time $s$ has propagated past the matter system (see Sec. \ref{Sec:input_output_relation}). Therefore $U^\dagger(t)a^\dagger_l(s)a_l(s)U(t) = a^\dagger_{l,\text{out}}(s)a_{l,\text{out}}(s)$.
\par
After the Heisenberg picture field operator has been expressed in terms of the output field $a_{l,\text{out}}(s)$ (or $a^\dagger_{l,\text{out}}(s)$), we apply the input-output relation (see Eqs. (\ref{Eq:input_output_relation}) and (\ref{Eq:input_output_relation_single_molecule})) to each $a_{l,\text{out}}(s)$. The Heisenberg picture system operators $L_H$ in the input-output relation are then expanded perturbatively.

\par
Applying the Heisenberg perturbative expansion (Eq. (\ref{Eq:A_H_perturbative_expansion})) to $L_{Hl}(s+\hat{\mathbf{k}}_l\cdot \mathbf{x}_j/c)$ in the many-molecule input-output relation (Eq. (\ref{Eq:input_output_relation})), we have
\begin{align}
\begin{split}
    a_{l,\text{out}}(s)=&a_l(s) + \sum_{j=1}^N L_{l,j}(s+\frac{\hat{\mathbf{k}}_l\cdot \mathbf{x}_j}{c})-i\sum_{j=1}^N\int^{s+\hat{\mathbf{k}}_l\cdot \mathbf{x}_j/c}_0 dt_1\,\Big[L_{l,j}(s+\frac{\hat{\mathbf{k}}_l\cdot \mathbf{x}_j}{c}), H(t_1)\Big]\\
    &+(-i)^2\sum_{j=1}^N \int^{s+\hat{\mathbf{k}}_l\cdot \mathbf{x}_j/c}_0 dt_2 \int^{t_2}_0 dt_1\,\Big[\Big[L_{l,j}(s+\frac{\hat{\mathbf{k}}_l\cdot \mathbf{x}_j}{c}),H(t_2)\Big],H(t_1)\Big]\\
    &+(-i)^3\sum_{j=1}^N \int^{s+\hat{\mathbf{k}}_l\cdot \mathbf{x}_j/c}_0 dt_3\int^{t_3}_0 dt_2 \int^{t_2}_0 dt_1\,\Big[\Big[\Big[L_{l,j}(s+\frac{\hat{\mathbf{k}}_l\cdot \mathbf{x}_j}{c}),H(t_3)\Big],H(t_2)\Big],H(t_1)\Big]+\cdots.
\label{Eq:IO_perturbative_expansion}
\end{split}
\end{align}
In the case of a single molecule located at $\mathbf{x}=\mathbf{0}$, the expansion of the input-output relation (Eq. (\ref{Eq:input_output_relation_single_molecule})) becomes
\begin{align}
\begin{split}
    a_{l,\text{out}}(s) =& a_l(s) + L_l(s) -i\int^s_0 dt_1\, \Big[L_l(s),H(t_1)\Big]\\
    & + (-i)^2 \int^s_0 dt_2\int^{t_2}_0 dt_1\,\Big[\Big[L_l(s),H(t_2)\Big],H(t_1)\Big]\\
    &+ (-i)^3 \int^s_0 dt_3\int^{t_3}_0 dt_2 \int^{t_2}_0 dt_1\,\Big[\Big[\Big[L_l(s),H(t_3)\Big],H(t_2)\Big], H(t_1)\Big] +\cdots.
\label{Eq:IO_perturbative_expansion_single_molecule}
\end{split}
\end{align}
\par
Eqs. (\ref{Eq:IO_perturbative_expansion}) and (\ref{Eq:IO_perturbative_expansion_single_molecule}) can be thought of as an expansion of the Heisenberg picture field operator $U^\dagger(t)a_l(s)U(t)$ using the second approach (see Eq. (\ref{Eq:A_s_t_perturbative_expansion})).
To see this explicitly for the case of a single molecule, notice that since the molecule is located at $\mathbf{x}=\mathbf{0}$, the condition of large enough time (i.e., $s_l(t,\mathbf{x}=\mathbf{0})>s$) becomes $t>s$. Using the single-molecule Hamiltonian (Eq. (\ref{Eq:H_int_main})), the first order expansion term in Eq. (\ref{Eq:A_s_t_perturbative_expansion}) then becomes 
\begin{align}
\begin{split}
    &\sum_{l'}\int^t_0 dt_1 [a_l(s), -a_{l'}(t_1)L_{l'}^\dagger(t_1)+a_{l'}^\dagger(t_1)L_{l'}(t_1)] \\
    &= \int^t_0 dt_1\,\delta(s-t_1) L_l(t_1) \\
    &= L_l(s).
\end{split}
\end{align}
This is exactly the second term in Eq. (\ref{Eq:IO_perturbative_expansion_single_molecule}). The higher order terms in Eq. (\ref{Eq:IO_perturbative_expansion_single_molecule}) also match with the higher order terms in the Heisenberg expansion of $U^\dagger(t)a_l(s)U(t)$. Therefore, we see that in the conventional perturbing-the-state approach (approach 1) and the perturbing-the-observable approach (approach 2), one needs to consider the output condition $s_l(t,\mathbf{x}_j)>s$ explicitly in the expansion. However, in the expansion of the input-output relation (approach 3), the output condition $s_l(t,\mathbf{x}_j)>s$ is automatically accounted for when applying the input-output relation.
\par
In the more general case where the field observable $A(s)$ is not equal to $a_l(s)$, but equal to a product of more than one field operators, the expansion of the Heisenberg-evolved field operator $A(s)$ is given by products of Eq. (\ref{Eq:IO_perturbative_expansion}) or (\ref{Eq:IO_perturbative_expansion_single_molecule}). In general, the resulting expansion takes a different form from that obtained through approach 2.

\subsection{Normal-ordered perturbative expansion of the optical signal}
\label{Sec:normal_ordered_perturbative_expansion_operator}
To evaluate the nested commutators in Eqs. (\ref{Eq:A_expectation_1}), (\ref{Eq:A_expectation_perturbation}), and (\ref{Eq:IO_perturbative_expansion}), we introduce the following commutator identity.
Given arbitrary field operators $A_1$ and $A_2$, and system operators $B_1$ and $B_2$,
\begin{subequations}
\begin{equation}
    [A_1 B_1, A_2 B_2] = A_1 A_2[B_1,B_2] + [A_1,A_2]B_2 B_1
\label{Eq:commutator_identity_1}
\end{equation}
\begin{equation}
    \qquad\qquad\qquad\,\,\, = A_2 A_1 [B_1,B_2] + [A_1,A_2] B_1 B_2.
\label{Eq:commutator_identity_2}
\end{equation}
\label{Eq:commutator_identities}
\end{subequations}
If the light-matter interaction is treated semi-classically (i.e., the field is described by real or complex numbers), then the complex numbers $A_1$ and $A_2$ commute with each other. Therefore, the first terms in both Eqs. (\ref{Eq:commutator_identity_1}) and (\ref{Eq:commutator_identity_2}) are identical, and the second terms in both Eqs. (\ref{Eq:commutator_identity_1}) and (\ref{Eq:commutator_identity_2}) are zero. We will call the first terms the ``matter commutator" terms and the second terms the ``field commutator" terms. 
\par
Having two different representations of the commutator in Eq. (\ref{Eq:commutator_identities}) allows us to write the perturbative expansion in terms of normal-ordered field operators. In the perturbative expansion, one of the two arguments in the commutator is the Hamiltonian $H(t)$ (see Eqs. (\ref{Eq:A_expectation_perturbation}) and (\ref{Eq:IO_perturbative_expansion})). Without loss of generality, we let the second argument of the commutator be $H(t)$, which is a sum of the absorption-type terms (i.e., taking the form $a(t)L^\dagger(t)$) and the emission-type terms (i.e., taking the form $a^\dagger(t)L(t)$). 
Therefore, the field operator $A_2$ is either an annihilation operator (taking the form $a(t)$) or a creation operator (taking the form $a^\dagger(t)$).
If $A_1$ is a normal-ordered field operator, then we can ensure that the commutator $[A_1B_1,A_2B_2]$ remains normal-ordered by using the identity Eq. (\ref{Eq:commutator_identity_1}) if $A_2$ is an annihilation operator, and using the identity Eq. (\ref{Eq:commutator_identity_2}) if $A_2$ is a creation operator. We will see how this works in an example in Sec. \ref{Sec:comparing_IO_to_conventional}.
\par
We note that, in the literature, a different form of commutator identity is used to perform perturbative expansions in quantum light spectroscopy \cite{Schlawin_2017_tutorial,Glenn_2015}. This identity is usually expressed in the superoperator form as
\begin{equation}
    (A_1B_1)_- = A_{1+}B_{1_-} + A_{1-}B_{1+},
\label{Eq:superoperator_commutator_identity}
\end{equation}
where $A_- X = [A,X]$ is the commutator superoperator, and $A_+ X = \{A,X\}/2$ is the anti-commutator superoperator, with a factor of $1/2$. Again, the first term on the right hand side is the matter commutator term, and the second term is the field commutator term, which vanishes in the semi-classical treatment of light-matter interaction. One can prove this identity by taking the arithmatic average of Eqs. (\ref{Eq:commutator_identity_1}) and (\ref{Eq:commutator_identity_2}). The average can be expressed as
\begin{equation}
    (A_1B_1)_- (A_2B_2) = A_{1+}B_{1_-} (A_2B_2) + A_{1-}B_{1+} (A_2B_2),
\end{equation}
which is just Eq. (\ref{Eq:superoperator_commutator_identity}) applied to the operator $A_2B_2$. The identity of Eq. (\ref{Eq:superoperator_commutator_identity}) takes a more symmetric form, but it does not provide a direct way to perform normal-ordered perturbative expansion.

\subsection{Order of magnitude estimates}
\label{Sec:order_of_magnitudes}
For the perturbative expansion to be a good description of the dynamics, the magnitude of the Hamiltonian $H(t)$ (i.e., the light-matter interaction strength) times the interaction time should be much less than $1$ (i.e., $||H||t \ll 1$). To provide a rough estimate of the magnitude of $H(t)$, we note that there are two types of operators in $H(t)$: the $a(t)$-type field operator (including $a^\dagger(t)$) and the $L(t)$-type matter operator (including $L^\dagger(t)$). 
From Sec. \ref{sec:interaction_pic_Hamiltonian}, we see that both $a(t)$ and $L(t)$ have the same physical dimension of $1/[\text{Time}]$, and therefore we can compare their relative magnitudes directly. Given a light pulse with a temporal width of $\sim \Delta$ and containing $m$ photons on average, we have $\int dt \langle a^\dagger(t)a(t)\rangle = m$. Therefore, to account for the effects from a light pulse in spatial mode $l$, we assign an order of magnitude of $\sqrt{m_l/\Delta_l}$ to each $a_l(t)$ and $a^\dagger_l(t)$, where $m_l$ and $\Delta_l$ are the number of photons and pulse width in the $l$-th spatial mode. In the case that the $l$-th spatial mode is in vacuum, we have $\int^T_0 dt \langle a_l(t)a_l^\dagger(t')\rangle = 1$. Therefore we assign an order of magnitude of $1/\sqrt{T}$ to $a_l(t)$ and $a^\dagger_l(t)$, where $T$ is the total amount of time considered. The order of magnitude of $L(t)$ is 
\begin{equation}
    L(t)\sim\sqrt{\frac{\omega_0^3 |\mathbf{d}|^2 \Delta\Omega  }{8\pi^2 \epsilon_0 c^3}},
\label{Eq:L_order_of_magnitude}
\end{equation}
as can be seen from Eq. (\ref{Eq:L_definition}). We note that the case of $\Delta\Omega=8\pi/3$ corresponds to maximal coupling between the molecule and the spatial mode, since Eq. (\ref{Eq:L_order_of_magnitude}) then becomes the square root of the total spontaneous emission rate into all spatial directions ($\Gamma=\omega_0^3|\mathbf{d}|^2/3\pi\epsilon_0 \hbar c^3$) \cite{Loudon_2000_book}. One can express the solid angle area $\Delta\Omega$ in terms of the cross section area $A$ of the beam, since the cross section area may be accessed more directly in experiments. Using the order of magnitude estimates $k_0^2\Delta\Omega\sim\sigma_\perp^2$ and $A\sim 1/\sigma_\perp^2$ (see Fig. (\ref{fig:paraxial})), we have $\Delta\Omega\sim \lambda_0^2/4\pi^2 A$, where the wavelength $\lambda_0$ is given by $\lambda_0=2\pi/k_0$. Re-writting Eq. (\ref{Eq:L_order_of_magnitude}) in terms of $A$, we have
\begin{equation}
    L(t)\sim \sqrt{\Gamma}\sqrt{\frac{3\lambda_0^2}{32\pi^3A}}.
\end{equation}
We now see that as the cross section area of the beam $A$ increases, the electric field felt by the molecule decreases, and the coupling between molecule and the photon field in the spatial mode decreases.
\par
Table (\ref{tab:order_of_magnitude_estimates}) lists the order of magnitude estimates of the quantities discussed above. Numerical values of these quantities are evaluated using typical values encountered in nonlinear spectroscopy with visible light pulses \cite{Arsenault2020vibronic,niedringhaus2018primary,ma2019both,bolzonello2021photocurrent}. We use the following values (unless otherwise noted): pulse width $\Delta=30$ fs, wavelength $\lambda_0=650$ nm, number of photon $m=3\times10^{10}$ (corresponding to a $10$ nJ pulse), total experimental time $T=10$ ps, beam cross section area $A=(100\,\mu\text{m})^2$, and dipole moment $|\mathbf{d}|=4$ Debye. 
\begin{table}[h]
    \centering
    \begin{tabular}{|c|c|c|}
        \hline
        operator & order of magnitude & typical numerical value (ps\textsuperscript{$-1/2$}) \\
        \hline
        $a(t)$ vacuum effect & $\sim 1/\sqrt{T}$ & $0.3$ \\
        \hline
        $a(t)$ pulse effect, & $\sim 1/\sqrt{\Delta}$ & $6$\\
        single photon& & \\
        \hline
        $a(t)$ pulse effect, & $\sim\sqrt{m/\Delta}$ & $1\times10^6$ \\
        $3\times10^{10}$ photons &&\\
        \hline
        $\sqrt{\Gamma}$ & $\sim\sqrt{\omega_0^3|\mathbf{d}|^2/3\pi\epsilon_0 \hbar c^3}$ & $4\times10^{-3}$ \\
        \hline
        $L(t)$ & $\sim \sqrt{\Gamma}\sqrt{3\lambda_0^2/32\pi^3A}$ & $2\times 10^{-6}$\\
        \hline
    \end{tabular}
    \caption{Order of magnitude estimates of various operators and parameters in the perturbative expansion.}
    \label{tab:order_of_magnitude_estimates}
\end{table}
\par
To confirm that the perturbative expansion is a good description under these numerical values, we consider for example a molecule interacting with a $30$ fs wide, $10$ nJ light pulse (corresponding to $\approx 3\times 10^{10}$ photons). In this case, $||H||\sim a(t)L(t)\sim 2 \,\text{ps}^{-1}$. Multiplying by the interaction time (i.e., the pulse width $\Delta$), we have $||H||t \sim 0.06 \ll 1$. Hence the perturbative expansion provides a good description for the dynamics. The magnitudes of $a(t)$ for single photon pulse effects and for the vacuum effect are even smaller, ensuring that these effects are well within the perturbative regime.

\subsection{Comparing the input-output formalism to the conventional perturbative approach --- photon flux}
\label{Sec:comparing_IO_to_conventional}
We now compare the perturbative input-output approach (approach 3 of Sec. \ref{Sec:IO_perturbation}) to the conventional perturbative approach. Since we have shown that the conventional perturbative approach in the interaction picture (approach 1 of Sec. \ref{sec:conventional_pert_interaction_pic}) produces identical expressions as approach 2 (Sec. \ref{Sec:conventional_perturbation_Heisenberg}) in the Heisenberg picture, we will refer to approach 2 as the conventional approach here. We consider the photon flux as the field observable and restrict ourselves to the case of one spatial mode interacting with one molecule located at the origin. We will collect the expansion terms in orders of $L(t)$, since this is typically the smallest parameter in the expansion (see Table (\ref{tab:order_of_magnitude_estimates})). 
\par
We will see that the only zeroth order ($\sim L(t)^0$) term is the input photon flux $a^\dagger_l(t)a_l(t)$. The terms that are first order in $L(t)$ are proportional to the expectation values $\langle L(t)\rangle$ and $\langle L^\dagger(t)\rangle$ in the matter degrees of freedom (assuming no initial correlation between the matter and the photon field). These expectation values are nonzero only when the initial state contains nonzero coherence between different excitation subspaces (e.g., $\langle e|\rho_{\text{sys}}(0)|g\rangle\neq 0$) because $L(t)$ is proportional to $|g\rangle\langle e|$ (see Eq. (\ref{Eq:L_definition})). We assume the initial matter state is the thermal state, which contains zero coherence between different excitation subspaces. Therefore, the first order terms ($\sim L(t)^1$) are zero. The lowest order non-trivial terms are the second order ($\sim L(t)^2$) terms. These terms correspond to spontaneous emission and linear response effects such as absorption and stimulated emission. 

\subsubsection{Input-output approach}
\label{sec:IO_approach_photon_flux}
In the input-output formalism, the output photon flux is given by $\langle a^\dagger_{l,\text{out}}(s)a_{l,\text{out}}(s)\rangle$, where the bracket $\langle\cdots\rangle$ denotes the expectation value with respect to the initial state $\rho_{\text{tot}}(0)$, which is assumed to be a product state between the matter system state and the field state. Expanding $a_{l,\text{out}}(s)$ and $a^\dagger_{l,\text{out}}(s)$ using Eq. (\ref{Eq:IO_perturbative_expansion_single_molecule}), we have
\begin{align}
\begin{split}
    \langle a^\dagger_{l,\text{out}}(s)a_{l,\text{out}}(s)\rangle&=\Big\langle \big(a^\dagger_l(s)+L^\dagger_{l}(s)-i\int^s_0 dt_1 [L^\dagger_l(s), H(t_1)]+\cdots\big)\\
    &\quad\qquad\big(a_l(s)+L_{l}(s)-i\int^s_0 dt_1 [L_l(s), H(t_1)]+\cdots\big)\Big\rangle \\
\label{Eq:IO_perturb_photon_flux}
\end{split}
\end{align}
Now we collect the expansion terms in orders of $L(t)$. The zeroth order term is $\langle a^\dagger_l(s)a_l(s)\rangle$, which is the input photon flux. The first order terms are $\langle a^\dagger_l(s)L_l(s)\rangle$ and its complex conjugate, $\langle L^\dagger_l(s) a_l(s)\rangle$. Due to the factorizable initial state, the first order term $\langle a^\dagger_l(s)L_l(s)\rangle$ can be factorized into $\langle a^\dagger_l(s)\rangle\langle L_l(s)\rangle$, which is equal to zero, as noted above.
\par
There are two types of second order terms. The first type of second order terms are products between a zeroth order term and a second order term, i.e., 
\begin{equation}
    \int^s_0 dt_1\,\Big\langle a_l^\dagger(s)\big[L_l(s),-a_l(t_1)L_l^\dagger(t_1)+a_l^\dagger(t_1)L_l(t_1)\big]\Big\rangle + \text{c.c.},
\end{equation}
where c.c. means the complex conjugate.
After expanding the commutator, we drop the term involving $\langle LL\rangle$ and $\langle L^\dagger L^\dagger \rangle$ because there is no initial coherence between different excitation subspaces. Keeping only the terms involving $\langle LL^\dagger\rangle$ and $\langle L^\dagger L\rangle$, we now have
\begin{equation}
    \int^s_0 dt_1\,\Big\langle a_l^\dagger(s)a_l(t_1)\Big\rangle\Big\langle L_l^\dagger(t_1)L_l(s)-L_l(s)L_l^\dagger(t_1)\Big\rangle + \text{c.c.}
\label{Eq:absorption_stimulated_emission_IO}
\end{equation}
The matter correlation function $\langle L_l(s)L_l^\dagger(t_1)\rangle$ correspond to the linear absorption process. 
The negative sign in front of $L_l(s)L_l^\dagger(t_1)$ is consistent with the fact that the absorption process reduces the output photon flux. The matter correlation function $\langle L_l^\dagger(t_1)L_l(s)\rangle$ correspond to the stimulated emission process. 
The positive sign in front of this term is consistent with the fact that stimulated emission increases the output photon flux.
\par
The second type of second order terms is $\langle L_l^\dagger(s)L_l(s)\rangle$, a product of two first order terms in Eq. (\ref{Eq:IO_perturb_photon_flux}).
This is the spontaneous emission rate into the $l$-th spatial mode. In the perturbative treatment of Eq. (\ref{Eq:IO_perturb_photon_flux}), the spontaneous emission rate is evaluated with respect to the initial matter state $\rho_{\text{sys}}(0)$. This is to be compared with the non-perturbative treatment of spontaneous emission in Eq. (\ref{Eq:photon_flux_vacuum}), where the exact spontaneous emission rate is evaluated with respect to the matter state $\rho_{\text{sys}}(s)$ at time $s$.

\subsubsection{Conventional perturbing-the-state approach}
Now we use the conventional approach to expand the photon flux directly, without first applying the input-output relation. 
\par
We let the molecule to be located at $\mathbf{x}=\mathbf{0}$, so the output photon flux takes the form
\begin{equation}
    a_{l,\text{out}}^\dagger(s) a_{l,\text{out}}(s) = U^\dagger(t)a_l^\dagger(s)a_l(s)U(t),
\end{equation}
where the time $t$ is large enough (i.e., $t>s$) so that the plane of the photon field $a_l(s)$ has propagated past the molecule at $\mathbf{x}=\mathbf{0}$. Applying the perturbative expansion of Eq. (\ref{Eq:A_expectation_perturbation}) to this Heisenberg-evolved operator to the second order, we have 
\begin{align}
\begin{split}
    \Big\langle a^\dagger_{l,\text{out}}(s)a_{l,\text{out}}(s)\Big\rangle =& \Big\langle a_l^\dagger(s)a_l(s)\Big\rangle -i \int^{t}_0 dt_1 \,\Big\langle[a_l^\dagger(s)a_l(s), H(t_1)]\Big\rangle \\
   & -\int^{t}_0 dt_2 \int^{t_2}_0 dt_1\,\Big\langle[[a_l^\dagger(s)a_l(s), H(t_2)] ,H(t_1)]\Big\rangle.
\end{split}
\end{align}
Note that the upper bound of the first integral is $t$, not $s$.
\par
\begin{figure}[h]
    \centering
    \includegraphics[scale=0.6]{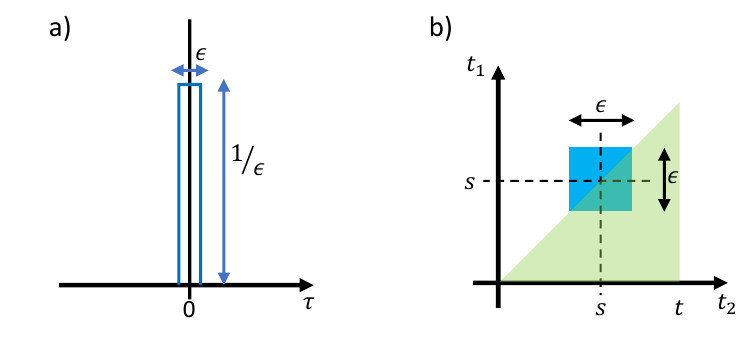}
    \caption{Evaluating the double integral in Eq. (\ref{Eq:conventional_spontaneous_emission_1}). (a) The delta function is treated as the $\epsilon\rightarrow 0$ limit of the square function with width $\epsilon$ and height $1/\epsilon$. (b) The product of the two delta functions in Eq. (\ref{Eq:conventional_spontaneous_emission_1}) is nonzero only in the blue square area. The nested integral over $t_2$ and $t_1$ is represented by the green area, and covers half of the delta function. Therefore, the double integral evaluates to $1/2$.}
    \label{fig:delta_function}
\end{figure}

The zeroth order term is the input photon flux. The first order term evaluates to 
\begin{align}
\begin{split}
    & \int^{t}_0 dt_1\,\delta(s-t_1) \Big\langle a_l(s)L_l^\dagger(t_1) + a_l^\dagger(s)L_l(t_1)\Big\rangle \\
    & = \Big\langle a_l(s)L_l^\dagger(s)\Big\rangle + \Big\langle a_l^\dagger(s)L_l(s)\Big\rangle,
\end{split}
\end{align}
which is the same as the first order term obtained in the input-output approach, and it is equal to $0$.
It is important to note that the delta function is integrated fully, since $t>s$. The second order term is
\begin{equation}
    \int^{t}_0 dt_2 \int^{t_2}_0 dt_1\,\delta(s-t_2)\Big\langle \Big[a_l(s)L_l^\dagger(t_2)+a_l^\dagger(s)L_l(t_2), -a_l(t_1)L_l^\dagger(t_1)+a_l^\dagger(t_1)L_l(t_1)\Big]\Big\rangle.
\label{Eq:conventional_second_order_expansion_1}
\end{equation}
To evaluate this commutator, we use Eq. (\ref{Eq:commutator_identities}) to ensure normal-ordering of the field operators. Following the arguments in Sec. \ref{sec:IO_approach_photon_flux}, we drop the terms involving the matter correlations $\langle LL\rangle$ or $\langle L^\dagger L^\dagger\rangle$. Eq. (\ref{Eq:conventional_second_order_expansion_1}) then becomes
\begin{align}
\begin{split}
    \int^{t}_0 dt_2 \int^{t_2}_0 dt_1\, \delta(s-t_2) \bigg( & \Big\langle a_l^\dagger(t_1)a_l(s)\Big\rangle\Big\langle[L_l^\dagger(t_2), L_l(t_1)]\Big\rangle \\
    & - \Big\langle a_l^\dagger(s)a_l(t_1)\Big\rangle\Big\langle[L_l(t_2), L_l^\dagger(t_1)]\Big\rangle\\
    & + \delta(s-t_1)\Big(\Big\langle L_l^\dagger(t_2)L_l(t_1)\Big\rangle + \Big\langle L_l^\dagger(t_1)L_l(t_2)\Big\rangle\Big)\bigg)
\label{Eq:conventional_second_order_expansion_2}
\end{split}
\end{align}
The first two terms in the big parenthesis are the matter commutator terms, and the last term involving the delta function $\delta(s-t_1)$ is the field commutator term. The sum of the first two terms is equal to Eq. (\ref{Eq:absorption_stimulated_emission_IO}) after one integrates over the delta function $\delta(s-t_2)$. The last term requires a careful treatment of the delta functions. First, we simplify the sum of the last two terms as
\begin{equation}
    2\big\langle L_l^\dagger(s) L_l(s)\big\rangle \int^{t}_0 dt_2 \int^{t_2}_0 dt_1\,\delta(s-t_2)\delta(s-t_1).
\label{Eq:conventional_spontaneous_emission_1}
\end{equation}
To integrate over the delta functions, we treat the delta function $\delta(\tau)$ as the $\epsilon\rightarrow 0$ limit of the rectangular function centered at $\tau=0$ with a width of $\epsilon$ and a height of $1/\epsilon$, so that the integral $\int d\tau\, \delta(\tau) = 1$ (see Fig. (\ref{fig:delta_function}a)). We choose the square function for the purpose of illustration. In fact, any symmetric function with integral equal to 1 that becomes infinitely narrow in the $\epsilon\rightarrow 0$ limit works as well. As shown in Fig. (\ref{fig:delta_function}b), the product of the two delta functions in Eq. (\ref{Eq:conventional_spontaneous_emission_1}) is nonzero only on a square region centered at $(t_1,t_2)=(s,s)$, having a width of $\epsilon$ (taking $\epsilon\rightarrow 0$ at the end). The integral of this entire square region is $1$. The double integral integrates over a triangular region in the $(t_1,t_2)$ plane, which cuts the square region in half. Therefore the double integral in Eq. (\ref{Eq:conventional_spontaneous_emission_1}) is equal to $1/2$, and the entire term of Eq. (\ref{Eq:conventional_spontaneous_emission_1}) becomes $\langle L^\dagger(s) L(s)\rangle$, equal to the second order spontaneous emission term derived using the input-output approach.

\subsubsection{Comparing the two approaches}
We see that in the conventional approach, extra care needs to be taken to ensure the upper bound $t$ of the first integral is larger than the retarded time argument $s$ in the field operator (e.g., $a^\dagger(s)a(s)$). If the position $\mathbf{x}$ of the molecule is not the origin $\mathbf{0}$, then we need to make sure that $s<s_l(t,\mathbf{x})=t-\hat{\mathbf{k}}_l\cdot\mathbf{x}/c$. In the literature, sometimes, this upper bound $t$ is taken to $\infty$, while sometimes, this upper bound is taken to be the retarded time variable $s$, and other times, the upper bound is not specified, leading to potential confusion as to whether the delta function should be integrated in full or in half.
On the other hand, in the input-output approach, one expands the Heisenberg-evolved system operator $L_H(s)$, and the upper bound of the first integral is $s$, the same as the time argument in the Heisenberg operator. The subtle integration bound in the conventional approach is taken care of by the input-output relation. Furthermore, we see that the derivation of terms like the spontaneous emission photon flux is much simpler using the input-output formalism than using the conventional approach.
\par
When the conventional approach is used with the normal-ordered perturbative expansion (described in Sec. \ref{Sec:normal_ordered_perturbative_expansion_operator}), one can obtain a perturbative expansion where all the field operators are normal-ordered. 
On the other hand, in the input-output approach, one expands the Heisenberg-evolved operators $L_H(s)$ in the signal expression (e.g., Eqs. (\ref{Eq:photon_flux_expansion}), (\ref{Eq:G_2_1}), and (\ref{Eq:HOM_IO})). The expansion of each $L_H(s)$ can be normal-ordered by applying Eq. (\ref{Eq:commutator_identities}). However, since the final signal expressions usually involve a product of multiple $L_H(s)$, the field operator in the expansion of the final signal becomes a product of normal-ordered operators, which is typically not normal-ordered.

\section{Conclusion}
We have developed a new input-output formalism for quantum light spectroscopy. The use of input-output relation provides physical intuition in analyzing photon field observables and simplifies the derivation for perturbative expansions. This formalism has helped us to discover an equivalence between a class of QLS and a class of classical light spectroscopy \cite{Ko_2023}. Further use of this input-output formulation could help the discovery of experimentally-feasible spectroscopic methods that exploit quantum properties of light.

\section*{Acknowledgements}
L.K. was supported by the Kavli Energy NanoScience Institute (ENSI) Philomathia graduate fellowship. This project was supported by the Photosynthetic Systems program of U.S. Department of Energy, Office of Science, Basic Energy Sciences, within the Division of Chemical Sciences, Geosciences, and Biosciences, under Award No. DESC0019728.

\newpage
\bibliographystyle{unsrt}
\typeout{}
\bibliography{references.bib}

\newpage
\appendix
\numberwithin{equation}{section}

\section{Expressing the 3-dimensional photon field in terms of 1-dimensional fields}
\label{app:photon_as_1D_fields}
To decompose the 3-dimensional field operators $a_{\mathbf{k},\lambda}$ into 1-dimensional field operators $a_l(\omega)$, we first re-write the multi-index $(\mathbf{k},\lambda)$ as $(|\mathbf{k}|,\Omega, \lambda)$, where $\Omega$ is the orientation of $\mathbf{k}$. One can further expand $\Omega$ into a polar angle $\theta$ and an azimuthal angle $\phi$, but we will use $\Omega$ for simplicity. The radial parameter $|\mathbf{k}|$ will be proportional to the 1-dimensional index $\omega$. We will perform a change of basis to represent $\Omega$ and $\lambda$ with a countably infinite set of basis functions $g_l(\Omega, \lambda)$, indexed by $l$.
\par
We find a complete and orthonormal set of functions $g_l(\Omega, \lambda)$ such that 
\begin{subequations}
    \begin{equation}
        \int d\Omega \sum_\lambda g^*_{l}(\Omega, \lambda)g_{l'}(\Omega, \lambda)=\delta_{l,l'}
    \label{Eq:g_orthonormal}
    \end{equation}
    \text{and}
    \begin{equation}
        \sum_{l=1}^{\infty} g^*_l(\Omega,\lambda)g_l(\Omega',\lambda')=\delta(\Omega-\Omega')\delta_{\lambda,\lambda'}.
    \label{Eq:g_complete}
    \end{equation}
\label{Eq:completeness_gl}
\end{subequations}
The integral over orientation $\int d\Omega$ can be expressed in terms of the polar angle $\theta$ and the azimuthal angle $\phi$ as $\int^\pi_0 d\theta \int^{2\pi}_0 d\phi \sin\theta$. The delta function $\delta(\Omega-\Omega')$ can be expressed as $\delta(\theta-\theta')\delta(\phi-\phi')/\sin\theta$.
\par
Let the frequency $\omega$ be equal to $c|\mathbf{k}|$, where $c$ is the speed of light. The 1-dimensional field $a_l(\omega)$ is defined as
\begin{equation}
    a_l(\omega) = \sqrt{\frac{\omega^2}{c^3}} \int d\Omega \sum_\lambda g_l(\Omega, \lambda) a_{|\mathbf{k}|, \Omega,\lambda}.
\label{Eq:a_l_definition}
\end{equation}
Using Eq. (\ref{Eq:completeness_gl}) and the delta-function identity $\delta(\mathbf{k}-\mathbf{k'})=\delta(|\mathbf{k}|-|\mathbf{k'}|)\delta(\Omega-\Omega')/|\mathbf{k}|^2$, one can show that the 1-dimensional field operators $a_l(\omega)$ satisfy the bosonic commutation relations: 
\begin{subequations}
\begin{equation}
    [a_l(\omega),a_{l'}(\omega')] = [a^\dagger_l(\omega),a^\dagger_{l'}(\omega')] = 0
\end{equation}
and
\begin{equation}
    [a_l(\omega),a^\dagger_{l'}(\omega')] = \delta(\omega-\omega')\delta_{l,l'}.
\end{equation}
\label{Eq:1D_boson_commutation_relations}
\end{subequations}
\par
Now, with the use of the completeness relation (i.e., Eq. (\ref{Eq:completeness_gl})), we can re-write $H_{\text{field}}$ (Eq. (\ref{Eq:H_field_1}) of the main text) in terms of the 1-dimensional fields $a_l(\omega)$ as
\begin{equation}
    H_{\text{field}} = \sum_{l=1}^\infty \int^\infty_0 d\omega\, \hbar\omega a^\dagger_l(\omega)a_l(\omega).
\label{Eq:H_field_2}
\end{equation}
To re-write the electric field operator in terms of $a_l(\omega)$, we first re-write the integral $\int d^3 k$ in Eq. (\ref{Eq:E_field_1}) of the main text as $\frac{1}{c}\int d\omega \int d\Omega$. Then
\begin{equation}
    \mathbf{E}(\mathbf{x}=\mathbf{0}) = \int^\infty_0 d\omega \sqrt{\frac{\hbar\omega^3}{16\pi^3c^3\epsilon_0}}\sqrt{\frac{\omega^2}{c^3}} \int d\Omega \sum_\lambda ia_{|\mathbf{k}|, \Omega,\lambda} \hat{\mathbf{e}}_{|\mathbf{k}|, \Omega,\lambda} + \text{h.c.}
\label{Eq:E_field_2}
\end{equation}
The abbreviation h.c. denotes the Hermitian conjugate of the previous term.
We suppose there is a finite number $l_\text{max}$ of real-valued mode functions $g_l(\Omega, \lambda)$ such that 
\begin{equation}
    \hat{\mathbf{e}}_{|\mathbf{k}|, \Omega,\lambda} = \sum_{l=1}^{l_{\text{max}}} C_l g_l(\Omega,\lambda) \hat{\mathbf{e}}_l 
\label{Eq:E_field_as_finite_1D_fields_condition}
\end{equation}
for some constants $C_l$ and unit vectors $\hat{\mathbf{e}}_l$. We will provide two specific examples of Eq. (\ref{Eq:E_field_as_finite_1D_fields_condition}) shortly, corresponding to the polarization mode decomposition and the small solid angle mode decomposition \cite{Ko_2022}.
Substituting Eq. (\ref{Eq:E_field_as_finite_1D_fields_condition}) into Eq. (\ref{Eq:E_field_2}) and using Eq. (\ref{Eq:a_l_definition}), we can now express $\mathbf{E}(\mathbf{x}=\mathbf{0})$ in terms of a finite number of 1-dimensional fields $a_l(\omega)$, i.e.,
\begin{equation}
    \mathbf{E}(\mathbf{x}=\mathbf{0}) = \sum_{l=1}^{l_{\text{max}}} C_l \int^\infty_0 d\omega\, \sqrt{\frac{\hbar\omega^3}{16\pi^3c^3\epsilon_0}} ia_l(\omega)\hat{\mathbf{e}}_l + \text{h.c.}
\label{Eq:E_field_finite_1D_app}
\end{equation}
This is the general form for writing $\mathbf{E}(\mathbf{x}=\mathbf{0})$ in terms of 1-dimensional fields $a_l(\omega)$. Depending on the choice of mode decomposition (i.e., Eqs. (\ref{Eq:completeness_gl}) and (\ref{Eq:E_field_as_finite_1D_fields_condition})), $C_l$ and $\hat{\mathbf{e}}_l$ will be different. We now present two examples of Eq. (\ref{Eq:E_field_as_finite_1D_fields_condition}), where the finitely many $g_l(\Omega,\lambda)$ are orthonormal to one another. From these finite number of orthonormal mode functions, it is then possible, in principle, to construct countably infinitely many more orthonormal $g_l(\Omega,\lambda)$ to form a complete set that satisfies Eq. (\ref{Eq:g_complete}). We will not construct the complete set of $g_l(\Omega,\lambda)$ explicitly, since we are only interested in expressing the electric field in terms of a finite number of modes. 
\par
In the polarization mode decomposition scheme, we decompose the electric field into three 1-dimensional fields, indexed by $l=x$, $y$, and $z$. The three 1-dimensional fields correspond to the three spatial components of the electric field. The mode functions are $g_l(\Omega,\lambda) = \sqrt{\frac{3}{8\pi}} \hat{\epsilon}_l\cdot \hat{\epsilon}_{\Omega,\lambda}$. The mode unit vectors are $\hat{\epsilon}_x = \hat{x}$, $\hat{\epsilon}_y = \hat{y}$, and $\hat{\epsilon}_z = \hat{z}$. The constants are $C_l=1$. One can then check directly that Eq. (\ref{Eq:E_field_as_finite_1D_fields_condition}) is satisfied and that $g_x$, $g_y$, and $g_z$ are indeed orthonormal to one another (i.e., satisfying Eq. (\ref{Eq:g_orthonormal})).
\par
In the small solid angle decomposition scheme, we partition all possible orientations $\Omega$ into $M$ number of small solid angle sections, indexed by $m$. Each small solid angle section can have two different polarizations, indexed by $p$ (to be distinguished from $\lambda$ in Eq. (\ref{Eq:E_field_as_finite_1D_fields_condition})). The index $l$ in Eq. (\ref{Eq:E_field_as_finite_1D_fields_condition}) is now a multi-index $(m,p)$, and $l_{\text{max}}=2M$. The mode functions $g_{m,p}$ are
\begin{equation}
    g_{m,p}(\Omega,\lambda)=
    \begin{cases}
        \frac{1}{\sqrt{\Delta\Omega_m}}\quad, \text{if $\Omega$ is in the solid angle section $m$ and $\lambda=p$} \\
        0\quad, \text{otherwise.}
    \end{cases}
\label{Eq:g_small_solid_angle_def}
\end{equation}
$\Delta\Omega_m$ is the area of the solid angle section $m$. The mode unit vector $\hat{\mathbf{e}}_{(m,p)}$ is the unit vector of the $p$-th polarization in the $m$-th small solid angle section. Note that we have assumed that the small solid angle sections are small enough such that we can define two constant polarization unit vectors within each small solid angle section. The constants are $C_{m,p} = \sqrt{\Delta\Omega_m}$. One can check that Eq. (\ref{Eq:E_field_as_finite_1D_fields_condition}) is satisfied and $g_{m,p}$ are orthonormal.

\section{Small solid angle mode in real space and time}
\label{app:small_solid_angle_real_space}
For convenience purposes, we copy Eq. (\ref{Eq:f_real_space_time}) here:
\begin{equation}
    \Tilde{f}(\mathbf{x},t)=\int d^3\mathbf{k} \, f(\mathbf{k}) e^{i\mathbf{k}\cdot \mathbf{x}} e^{-ic|\mathbf{k}|t}.
\label{Eq:f_real_space_time_app}
\end{equation}
$f(\mathbf{k})$ is a function that has significant amplitude only in a small region $R$ in $\mathbf{k}$-space. $R$ is centered at $\mathbf{k}_l$. It has a transversal width of $\sigma_\perp$ and a longitudinal width of $\sigma_\parallel$. 
Without loss of generality, we let $\mathbf{k}_l$ to be located on the z-axis. We assume $f(\mathbf{k})$ is slowly varying such that within the small region, $|\partial  f(\mathbf{k})/\partial k_x|$ and $|\partial  f(\mathbf{k})/\partial k_y|$ is at most on the order of $1/\sigma_\perp$, and that $|\partial  f(\mathbf{k})/\partial k_z|$ is at most on the order of $1/\sigma_\parallel$.
\par
Without assuming a specific functional form for $f(\mathbf{k})$, we can already understand the behavior of $\Tilde{f}(\mathbf{x},t=0)$ and $\Tilde{f}(\mathbf{x},t\rightarrow \infty)$. 
When $t=0$,
\begin{equation}
    \Tilde{f}(\mathbf{x},0) = \int d^3\mathbf{k}\,f(\mathbf{k})e^{i\mathbf{k}\cdot\mathbf{x}}
\end{equation}
is simply the Fourier transform of $f(\mathbf{k})$. Since $f(\mathbf{k})$ is a slowly-varying function having a transversal span of $\sim \sigma_\perp$, $\Tilde{f}(\mathbf{x},0)$ has significant amplitude only within a transversal cross sectional area of $\sim 1/\sigma_\perp^2$.
When $t\rightarrow\infty$, the the complex exponent in the integrand of Eq. (\ref{Eq:f_real_space_time_app}) tends to be fast-varying with respect to $|\mathbf{k}|$. Therefore the integrand tends to be highly oscillatory in the $k_z$ direction, making the integral tend to $0$. For $\Tilde{f}(\mathbf{x},t)$ to be nonzero, the exponent
\begin{equation}
    i\phi = i(\mathbf{k}\cdot\mathbf{x}-c|\mathbf{k}|t) 
\end{equation}
in Eq. (\ref{Eq:f_real_space_time_app})
needs to be stationary with respect to $\mathbf{k}$, for some $\mathbf{k}$ in the small region $R$. The stationary condition is obtained by setting the derivative $\nabla_\mathbf{k}\phi$ to be zero. After calculating the derivative, we have the stationary condition as
\begin{equation}
    \mathbf{x} = \hat{\mathbf{k}}ct,
\label{Eq:stationary_condition}
\end{equation}
where $\hat{\mathbf{k}}=\mathbf{k}/|\mathbf{k}|$ is the unit vector in the direction of $\mathbf{k}$.
In order for the condition of (\ref{Eq:stationary_condition}) to hold for some $\mathbf{k}$ in $R$, $\mathbf{x}$ needs to have a magnitude of $ct$ and a direction that lies inside the small solid angle section. Therefore, $\Tilde{f}(\mathbf{x},t\rightarrow\infty)$ has significant amplitude only when $\mathbf{x}$ lies inside the cone of the small solid angle.

\par
To give a specific example, let $f(\mathbf{k})$ be the Gaussian function
\begin{equation}
    f(\mathbf{k}) = \exp\Big(-\frac{k_x^2+k_y^2}{2\sigma_\perp^2} - \frac{\zeta^2}{2\sigma_\parallel^2}\Big),
\end{equation}
where $\zeta = k_z-k_0$ and $k_0 = |\mathbf{k}_l|$.
To solve the integral in Eq. (\ref{Eq:f_real_space_time_app}) approximately, we approximate $|\mathbf{k}|=\sqrt{k_x^2+k_y^2+k_z^2}$ in the exponent by its Taylor expansion around $k_0$. Keeping only the lowest order terms in $k_x$, $k_y$, and $\zeta$,
\begin{align}
\begin{split}
    |\mathbf{k}| = k_0 + \zeta + \frac{k_x^2}{2k_0} + \frac{k_y^2}{2k_0}.
\end{split}
\end{align}
Now the integral in Eq. (\ref{Eq:f_real_space_time_app}) becomes
\begin{align}
\begin{split}
    \Tilde{f}(x,y,z,t) = e^{ik_0 z - ick_0 t} &\int dk_x \, \exp \Big(- k_x^2\big(\frac{1}{2\sigma_\perp^2} + \frac{ict}{2k_0}\big) + ik_x x\Big) \\
    &\int dk_y\, \exp \Big(- k_y^2\big(\frac{1}{2\sigma_\perp^2} + \frac{ict}{2k_0}\big) + ik_y y\Big) \\
    &\int d\zeta \, \exp \Big(- \frac{\zeta^2}{2\sigma_\perp^2} + i\zeta (z-ct)\Big).
\end{split}
\end{align}
After performing the Gaussian integrals, we see that
\begin{align}
\begin{split}
    |\Tilde{f}(x,y,z,t)| = A \exp \Big( -\frac{x^2+y^2}{2w_\perp^2} -\frac{(z-ct)^2\sigma_\parallel^2}{2} \Big),
\end{split}
\end{align}
where the transversal width $w_\perp$ in real space is
\begin{equation}
    w_\perp = \frac{1}{\sigma_\perp}\sqrt{1+\Big(\frac{ct\sigma_\perp^2}{k_0}\Big)^2},
\end{equation}
and the position-independent constant $A$ is
\begin{equation}
    A = \sqrt{\frac{2\sigma_\parallel^2 \pi^3}{\Big( \frac{1}{2\sigma_\perp^2} \Big)^2 + \Big( \frac{ct}{2k_0} \Big)^2}}.
\end{equation}
In the longitudinal direction, $\Tilde{f}(x,y,z,t)$ behaves as a wavepacket traveling at the speed of light due to the Gaussian factor centered at $z-ct$. The transverse width is time-dependent, and it is minimal at $t=0$.
When $t=0$, $w_\perp = 1/\sigma_\perp$, so the cross section area is on the order of $1/\sigma_\perp^2$. At long enough time such that $t\gg \frac{k_0}{c\sigma_\perp^2}$, $w_\perp = \frac{ct\sigma_\perp}{k_0}$. 
At long times, the pulse is centered around $z=ct$ in the $z$-direction, so the divergence angle of the small solid angle mode in real space and time is given by $\Delta\theta = w_\perp/ct = \sigma_\perp/k_0$. Note that this divergence angle in real space is the same as the angular spread of the small region $R$ in $\mathbf{k}$-space, given by $\sigma_\perp/k_0$.

\section{Classical input-output relation}
\label{app:classical_IO}
Our starting point for the derivation of the classical input-output relation is the inhomogeneous wave equation for the electric field
\begin{equation}
    \nabla^2 \mathbf{E} - \frac{1}{c^2}\frac{\partial^2\mathbf{E}}{\partial t^2} = \frac{1}{\epsilon_0 c^2}\frac{\partial^2 \mathbf{P}}{\partial t^2}
\label{Eq:E_wave_equation}
\end{equation}
derived from the macroscopic Maxwell's equations \cite{Boyd_nonlinear}, where $\mathbf{E}$ is the electric field and $\mathbf{P}$ is the polarization field, or the density of the dipole moment. Let $\mathbf{E}$ and $\mathbf{P}$ take the form
\begin{subequations}
\begin{equation}
    \mathbf{E}(\mathbf{r},t) = \Tilde{\mathbf{E}}(z,t)e^{ikz-i\omega t} + \text{c.c.}
\end{equation}
\begin{equation}
    \mathbf{P}(\mathbf{r},t) = \Tilde{\mathbf{P}}(z,t)e^{ikz-i\omega t} + \text{c.c.},
\end{equation}
\label{Eq:E_and_P_form}
\end{subequations}
where $\omega=ck$ is the carrier wave frequency, and $z$ is in the direction of the field propagation. The notation c.c. stands for complex conjugate. In Eq. (\ref{Eq:E_and_P_form}) we have assumed that $\mathbf{E}(\mathbf{r},t)$ and $\mathbf{P}(\mathbf{r},t)$ change very slowly in the transverse $x$ and $y$ directions, so that the dependence on $x$ and $y$ is ignored.
We further assume that $\Tilde{\mathbf{E}}$ and $\Tilde{\mathbf{P}}$ are slowly varying envelope functions such that
\begin{subequations}
\begin{equation}
    \Big|\frac{\partial^2\Tilde{\mathbf{E}}}{\partial t^2}\Big| \ll \omega\Big|\frac{\partial\Tilde{\mathbf{E}}}{\partial t}\Big| \ll \omega^2\Big|\Tilde{\mathbf{E}}\Big| \quad \text{and} \quad \Big|\frac{\partial^2\Tilde{\mathbf{E}}}{\partial z^2}\Big| \ll k\Big|\frac{\partial\Tilde{\mathbf{E}}}{\partial z}\Big| \ll k^2\Big|\Tilde{\mathbf{E}}\Big|
\end{equation}
and
\begin{equation}
    \Big|\frac{\partial^2\Tilde{\mathbf{P}}}{\partial t^2}\Big| \ll \omega\Big|\frac{\partial\Tilde{\mathbf{P}}}{\partial t}\Big| \ll \omega^2\Big|\Tilde{\mathbf{P}}\Big| \quad \text{and} \quad \Big|\frac{\partial^2\Tilde{\mathbf{P}}}{\partial z^2}\Big| \ll k\Big|\frac{\partial\Tilde{\mathbf{P}}}{\partial z}\Big| \ll k^2\Big|\Tilde{\mathbf{P}}\Big|.
\end{equation}
\label{Eq:SVEA}
\end{subequations}
Substituting Eq. (\ref{Eq:E_and_P_form}) into the second order equation of Eq. (\ref{Eq:E_wave_equation}) and applying the slowly varying envelope approximation (Eq. (\ref{Eq:SVEA})), we obtain a first order equation
\begin{equation}
    \frac{\partial \Tilde{\mathbf{E}}}{\partial z} + \frac{1}{c}\frac{\partial\Tilde{\mathbf{E}}}{\partial t} = \frac{ik}{2\epsilon_0}\Tilde{\mathbf{P}}.
\label{Eq:first_order_wave_eqn_1}
\end{equation}
\par
Defining a retarded time
\begin{equation}
    s = t-\frac{z}{c}
\end{equation}
and changing the variables in Eq. (\ref{Eq:first_order_wave_eqn_1}) from $(z,t)$ to $(s,t)$, we can reduce the partial differential equation in Eq. (\ref{Eq:first_order_wave_eqn_1}) to an ordinary differential equation of the variable $t$, for each fixed $s$, i.e.,
\begin{equation}
    \frac{\partial }{\partial t} \Tilde{\mathbf{E}}'(s, t) =  \frac{ikc}{2\epsilon_0}\Tilde{\mathbf{P}}'(s, t).
\label{Eq:first_order_wave_eqn_2}
\end{equation}
To be clear on the variables, we have defined $\Tilde{\mathbf{E}}'(s,t) = \Tilde{\mathbf{E}}(z,t)$ and $\Tilde{\mathbf{P}}'(s,t) = \Tilde{\mathbf{P}}(z,t)$.
Eq. (\ref{Eq:first_order_wave_eqn_2}) can be solved as
\begin{equation}
    \Tilde{\mathbf{E}}'(s,t) = \Tilde{\mathbf{E}}'(s,0) + \frac{ikc}{2\epsilon_0} \int^t_{0} d\tau\, \Tilde{\mathbf{P}}'(s, \tau).
\label{Eq:classical_IO_relation_1}
\end{equation}
Let the matter sample be located between $z=-\epsilon$ and $z=\epsilon$. Then the polarization field $\Tilde{\mathbf{P}}(z,t)$ is nonzero only when $|z| < \epsilon$, and $\Tilde{\mathbf{P}}'(s,t)$ is nonzero only when $|t-s|<\epsilon/c$. The input field $\mathbf{F}_{\text{in}}(s)$ is defined as $\Tilde{\mathbf{E}}'(s,0)$, where $s > \epsilon/c$, so that the field is upstream of the matter sample. The output field $\mathbf{F}_{\text{out}}(s)$ is defined as $=\Tilde{\mathbf{E}}'(s,t)$, where $t>s+\epsilon/c$, so that the field is downstream of the matter sample. Now, we can re-write Eq. (\ref{Eq:classical_IO_relation_1}) as
\begin{equation}
    \mathbf{F}_{\text{out}}(s) = \mathbf{F}_{\text{in}}(s) + \frac{ikc}{2\epsilon_0} \int^{s+\epsilon/c}_{s-\epsilon/c} d\tau \, \Tilde{\mathbf{P}}'(s,\tau).
\label{Eq:classical_IO_relation_2}
\end{equation}
This is the classical input-output relation. Note that, given a fixed $s$, $\Tilde{\mathbf{P}}'(s,\tau)$ is nonzero only when $\tau\in (s-\epsilon/c, s+\epsilon/c)$. By the definitions of the input and output fields, $(s-\epsilon/c, s+\epsilon/c)$ is always contained inside $(0,t)$.
\par
To turn this into a form that is more similar to Eq. (\ref{Eq:input_output_relation}), we convert $\Tilde{\mathbf{P}}'
$ into $\Tilde{\mathbf{P}}$ in the integral inside Eq. (\ref{Eq:classical_IO_relation_2}), and re-write the integral as
\begin{equation}
    \int^{s+\epsilon/c}_{s-\epsilon/c} d\tau\, \Tilde{\mathbf{P}}(c(\tau-s), \tau).
\end{equation}
Changing the integration variable from $\tau$ to $z=c(\tau-s)$, the integral becomes
\begin{equation}
    \frac{1}{c}\int^\epsilon_{-\epsilon} dz \, \Tilde{\mathbf{P}}(z,s+z/c).
\end{equation}
We can now write the classical input-output relation as
\begin{equation}
    \mathbf{F}_{\text{out}}(s) = \mathbf{F}_{\text{in}}(s) + \frac{ik}{2\epsilon_0}\int^\epsilon_{-\epsilon} dz \, \Tilde{\mathbf{P}}(z,s+z/c).
\end{equation}
Comparing this to the quantum input-output relation (Eq. (\ref{Eq:input_output_relation})), the time variable $s+z/c$ of the polarization $\Tilde{\mathbf{P}}$ here corresponds to the time variable $s+\frac{\hat{\mathbf{k}}_l\cdot \mathbf{x}_j}{c}$ of the dipole operator in Eq. (\ref{Eq:input_output_relation}). The integral over $z$ here corresponds to the sum over molecules in Eq. (\ref{Eq:input_output_relation}). The factor of $i$ in the classical expression comes from the correspondence $ia(s)\leftrightarrow \mathbf{E}$ (ignoring a real-valued constant factor) between the field operator $a(s)$ and the classical electric field $\mathbf{E}$.

\section{Fermi's golden rule rate for spontaneous emission photon flux}
\label{app:spontaneous_emission_Fermi_rate}
Consider a molecule with two electronic states, a ground state $|g\rangle$ and an excited state $|e\rangle$. The molecule may contain nuclear degrees of freedom, which will be implicit in the analysis. Let the initial electronic state of the molecule be in the excited state. Then the golden rule expression for the decay rate from $|e\rangle$ into $|g\rangle$ can be written in terms of the correlation function \cite{Nitzan_book}, i.e., 
\begin{equation}
    \text{excited state decay rate} = \int^\infty_{-\infty} d\tau\,\langle H(\tau)H(0)\rangle.
\label{Eq:Golden_rule_rate_1}
\end{equation}
Here, the expectation value is evaluated with respect to $|e\rangle|\text{vac}\rangle$, the electronically excited state of the molecule times the vacuum state of the photon field. Substituting the Hamiltonian (Eq. (\ref{Eq:H_int_main})) into the golden rule expression, we find the excited state decay rate is equal to
\begin{equation}
    \text{excited state decay rate} = \sum_{l=1}^{2M} \langle e |L^\dagger_l L_l |e\rangle,
\end{equation}
meaning that the total decay rate of the excited state is the sum of the spontaneous emission rates into each of the spatial modes. Given a general system state $\rho_{\text{sys}}(t)$ that may not be the excited state $|e\rangle$, the spontaneous emission photon flux in the $l$-th spatial mode is equal to the excited state population $\text{Tr}(| e\rangle\langle e| \rho_{\text{sys}}(t))$ times the emission rate into the $l$-th spatial mode $\langle e|L^\dagger_l L_l |e\rangle$. This product can be shown to be equivalent to Eq. (\ref{Eq:photon_flux_vacuum}) by noting that
\begin{align}
\begin{split}
    &\quad \text{Tr}\big(|e\rangle\langle e|\rho_{\text{sys}}(t)\big) \langle e|L_l^\dagger L_l|e\rangle \\
    &=\text{Tr}\big(|e\rangle\langle e|L_l^\dagger L_l|e\rangle\langle e|\rho_{\text{sys}}(t)\big)\\
    &=\text{Tr}\big(L_l^\dagger L_l\rho_{\text{sys}}(t)\big).
\end{split}
\end{align}
The last equality is obtained by noting that in our two-level system, $L^\dagger_l L_l$ is proportional to $|e\rangle\langle e|$.

\section{Perturbative expansion of Heisenberg-evolved operators}
\label{app:Heisenberg_perturbation}

Following the notation in Sec. \ref{Sec:perturbative_expansion_of_Heisenberg_operator}, we define a time-evolution superoperator $\mathcal{U}(t)$ that evolves an operator in time as
\begin{equation}
    \mathcal{U}(t) = U^\dagger(t)\bullet U(t),
\label{Eq:time_evol_super_def}
\end{equation}
where the dot $\bullet$ notation means that the effect of the superoperator acting on a general operator $X$ is to substitute the operator $X$ into the dot $\bullet$, i.e., $\mathcal{U}(t) X = U^\dagger(t) X U(t)$.
Note that a Heisenberg picture operator $A_H(t)$ is related to the interaction picture operator $A(t)$ by
\begin{equation}
    A_H(t) = \mathcal{U}(t)A(t)
\label{Eq:Heisenberg_interaction_pic_operator}.
\end{equation}
Our task of expanding $A_H(t)$ from $A(t)$ now becomes expanding the superoperator $\mathcal{U}(t)$.
Just as $U(t)$ is a linear operator on the vector space of quantum states, the superoperator $\mathcal{U}(t)$ is also a linear operator on a larger vector space of operators. 

\par
Taking the time derivative of Eq. (\ref{Eq:time_evol_super_def}) using the Schrodinger equation (Eq. (\ref{Eq:Schrodinger_eqn_1})), we obtain
\begin{align}
\begin{split}
    \frac{d}{dt} \mathcal{U}(t) & = -iU^\dagger(t)[\bullet, H(t) ]U(t) \\
    & = -i[U^\dagger(t)\bullet U(t), H_H(t) ],
\label{Eq:forward_Heisenberg}
\end{split}
\end{align}
where $H_H(t) = U^\dagger(t)H(t)U(t)$ is the Hamiltonian in the Heisenberg picture. We have written the time derivative in two ways. The first line involves the interaction picture Hamiltonian $H(t)$, and the second line involves the Heisenberg picture Hamiltonian $H_H(t)$.
Defining the superoperator $\mathcal{L}[X]$, where $X$ is a general operator, as
\begin{equation}
    \mathcal{L}[X] = -i[\bullet, X],
\end{equation}
we can re-express Eq. (\ref{Eq:forward_Heisenberg}) in superoperators in two ways.
The second line in Eq. (\ref{Eq:forward_Heisenberg}) can be written as
\begin{equation}
    \frac{d}{dt}\mathcal{U}(t) = \mathcal{L}[H_H(t)]\mathcal{U}(t),
\label{Eq:time_ordered_1}
\end{equation}
while the first line in Eq. (\ref{Eq:forward_Heisenberg}) can be written as
\begin{equation}
    \frac{d}{dt}\mathcal{U}(t) = \mathcal{U}(t)\mathcal{L}[H(t)].
\label{Eq:anti_time_ordered_1}
\end{equation}
\par
Solving Eq. (\ref{Eq:time_ordered_1}) iteratively from the initial condition $\mathcal{U}(0)=1$, we have
\begin{align}
\begin{split}
    \mathcal{U}(t) = & 1 + \int^t_{0} dt_1 \,\mathcal{L}[H_H(t_1)] \\
    & + \int^t_{0} dt_2 \int^{t_2}_{0} dt_1\, \mathcal{L}[H_H(t_2)]\mathcal{L}[H_H(t_1)] \\
    & + \int^t_{0} dt_3 \int^{t_3}_{0} dt_2\int^{t_2}_{0} dt_1\, \mathcal{L}[H_H(t_3)]\mathcal{L}[H_H(t_2)]\mathcal{L}[H_H(t_1)] + \cdots.
\label{Eq:time_ordered_2}
\end{split}
\end{align}
This result can be expressed compactly as 
\begin{equation}
    \mathcal{U}(t) = \mathcal{T}_+ \exp \bigg(\int^t_{0}d\tau\,\mathcal{L}[H_H(\tau)]\bigg),
\label{Eq:time_ordered_3}
\end{equation}
where the forward time-ordering $\mathcal{T}_+$ orders the superoperators $\mathcal{L}$ from right to left in the order of the smallest to the largest time variable.
Substituting this expansion into Eq. (\ref{Eq:Heisenberg_interaction_pic_operator}), we obtain the time-ordered expansion for $A_H(t)$ as
\begin{align}
\begin{split}
    A_H(t)=& A(t)-i\int^t_{0}dt_1\,[A(t),H_H(t_1)] \\
    &+(-i)^2\int^t_{0}dt_2\int^{t_2}_{0}dt_1\,[[A(t),H_H(t_1)],H_H(t_2)]\\
    &+(-i)^3\int^t_{0}dt_3\int^{t_3}_{0}dt_2\int^{t_2}_{0}dt_1\,[[[A(t),H_H(t_1)],H_H(t_2)],H_H(t_3)]+\cdots.
\end{split}
\end{align}
This is the expansion of Eq. (\ref{A_H_expansion_1_time_ordered}), copied here for convenience.
\par
On the other hand, solving Eq. (\ref{Eq:anti_time_ordered_1}) iteratively from the initial condition $\mathcal{U}(0)=1$, we have
\begin{align}
\begin{split}
    \mathcal{U}(t) = & 1+\int^t_{0} dt_1 \,\mathcal{L}[H(t_1)] \\
    & + \int^t_{0} dt_2 \int^{t_2}_{0} dt_1\, \mathcal{L}[H(t_1)]\mathcal{L}[H(t_2)] \\
    & + \int^t_{0} dt_3 \int^{t_3}_{0} dt_2\int^{t_2}_{0} dt_1\, \mathcal{L}[H(t_1)]\mathcal{L}[H(t_2)]\mathcal{L}[H(t_3)] + \cdots.
\label{Eq:anti_time_ordered_2}
\end{split}
\end{align}
This result can be expressed compactly as 
\begin{equation}
        \mathcal{U}(t) = \mathcal{T}_-\exp \bigg( \int^{t}_{0} d\tau\, \mathcal{L}[H(\tau)] \bigg),
\label{Eq:anti_time_ordered_3}
\end{equation}
where the backward time-ordering $\mathcal{T}_-$ orders the superoperators $\mathcal{L}$ from right to left in the order of the largest to the smallest time variable.
Substituting this expansion into Eq. (\ref{Eq:Heisenberg_interaction_pic_operator}), we obtain the anti-time-ordered expansion for $A_H(t)$ as
\begin{align}
\begin{split}
    A_H(t)=& A(t)-i\int^t_{0}dt_1\,[A(t),H(t_1)] \\
    &+(-i)^2\int^t_{0}dt_2\int^{t_2}_{0}dt_1\,[[A(t),H(t_2)],H(t_1)]\\
    &+(-i)^3\int^t_{0}dt_3\int^{t_3}_{0}dt_2\int^{t_2}_{0}dt_1\,[[[A(t),H(t_3)],H(t_2)],H(t_1)]+\cdots.
\end{split}
\end{align}
This is the expansion of Eq. (\ref{Eq:A_H_perturbative_expansion}), copied here for convenience.
\par
We have now obtained two different expansions of the Heisenberg picture operator $A_H(t)$. Comparing Eq. (\ref{Eq:time_ordered_3}) to Eq. (\ref{Eq:anti_time_ordered_3}), we find a surprisingly elegant superoperator identity:
\begin{equation}
    \mathcal{T}_+ \exp \bigg(\int^t_{0}d\tau\,\mathcal{L}[H_H(\tau)]\bigg) = \mathcal{T}_-\exp \bigg( \int^{t}_{0} d\tau\, \mathcal{L}[H(\tau)] \bigg).
\end{equation}

\end{document}